# EARLY WARNING ANALYSIS FOR SOCIAL DIFFUSION EVENTS

Richard Colbaugh[*]     Kristin Glass[†]

[*]Sandia National Laboratories, Albuquerque, NM USA
[†]New Mexico Institute of Mining and Technology, Socorro, NM USA

**Abstract**

There is considerable interest in developing predictive capabilities for social diffusion processes, for instance to permit early identification of emerging contentious situations, rapid detection of disease outbreaks, or accurate forecasting of the ultimate reach of potentially "viral" ideas or behaviors. This paper proposes a new approach to this predictive analytics problem, in which analysis of meso-scale network dynamics is leveraged to generate useful predictions for complex social phenomena. We begin by deriving a stochastic hybrid dynamical systems (S-HDS) model for diffusion processes taking place over social networks with realistic topologies; this modeling approach is inspired by recent work in biology demonstrating that S-HDS offer a useful mathematical formalism with which to represent complex, multi-scale biological network dynamics. We then perform formal stochastic reachability analysis with this S-HDS model and conclude that the outcomes of social diffusion processes may depend crucially upon the way the early dynamics of the process interacts with the underlying network's *community structure* and *core-periphery structure*. This theoretical finding provides the foundations for developing a machine learning algorithm that enables accurate early warning analysis for social diffusion events. The utility of the warning algorithm, and the power of network-based predictive metrics, are demonstrated through an empirical investigation of the propagation of political "memes" over social media networks. Additionally, we illustrate the potential of the approach for security informatics applications through case studies involving early warning analysis of large-scale protests events and politically-motivated cyber attacks.

## 1. Introduction

Understanding the way information, behaviors, innovations, and diseases propagate over social networks is of great importance in a wide variety of domains [e.g., 1-4], including national security [e.g., 5-13]. Of particular interest are *predictive* capabilities for social diffusion, for instance to enable early warning concerning the emergence of a violent conflict or outbreak of an epidemic. As a consequence, vast resources are devoted to the task of predicting the outcomes of diffusion processes, but the quality of such predictions is often poor. It is tempting to conclude that the problem is one of insufficient information. Clearly diffusion phenomena which "go viral" are qualitatively different from those that don't or they wouldn't be so dominant, the conventional wisdom goes, so in order to make good predictions we must collect enough data to allow these crucial differences to be identified.

Recent research calls into question this intuitively plausible premise and, indeed, indicates that intuition can be an unreliable guide to constructing successful prediction methods. For example, studies of the predictability of popular culture indicate that the *intrinsic* attributes commonly believed to be important when assessing the likelihood of adoption of cultural products, such as the quality of the product itself, do not possess much predictive power [14-16]. This research offers evidence that, when individuals are influenced by the actions of others, it may not be possible to obtain reliable predictions using methods which focus on intrinsics alone; instead, it may be necessary to incorporate aspects of *social influence* into the prediction process. Very recently a handful of investigations have shown the value of considering even simple and indirect measures of social influence, such as early social media "buzz", when forming predictions. This work has produced useful prediction algorithms for an array of social phenomena, including markets [16-21], political and social movements [17,22], mobilization and protest behavior [23,24], epidemics [17,25], social media dynamics [26,27], and the evolution of cyber threats [28].

Recognizing the importance of accounting for social influence, this paper proposes a predictive methodology which explicitly considers the way individuals influence one another *through their social networks*. It is expected that prediction algorithms which are based, in part, on network dynamics metrics will outperform existing methods and be applicable to a wider range of diffusion systems. We begin by developing a stochastic hybrid dynamical systems (S-HDS) model for diffusion processes taking place over social networks with realistic topologies. This modeling approach is inspired by recent work in biology demonstrating that S-HDS offer a useful mathematical formalism with which to represent multi-scale biological network dynamics [29-33]. An S-HDS is a feedback interconnection of a discrete-state stochastic process, such as a Markov chain, with a family of continuous-state stochastic dynamical systems [34]. Combining discrete and continuous dynamics in this way provides a rigorous, expressive, and computationally-tractable framework for modeling the dynamics of the complex, highly-evolved networks that are ubiquitous in biological systems [35], and we show in this paper that the S-HDS framework is also well-suited to the task of modeling the network dynamics which underlie social diffusion.

With the S-HDS model in hand, we then perform formal stochastic reachability analysis and conclude that the outcomes of social diffusion processes may depend crucially upon the way the early dynamics of the process propagates with respect to the underlying network's 1.) *community structure*, that is, densely connected groupings of individuals which have only relatively few links to other groups [36], and 2.) *core-periphery structure*, reflecting the presence of a small group of "core" individuals that are densely connected to each other and are also close to the remainder of the network [36]. This theoretical finding leads to the identification of novel metrics for the community and core-periphery dynamics which should be useful early indicators of which diffusion events will propagate widely, ultimately affecting a substantial portion of the population of interest, and which will not. Prediction is accomplished with a machine learning algorithm [37] which is based, in part, on these network dynamics metrics.

The paper makes three main contributions. First, we present a new S-HDS-based framework for modeling social diffusion on networks of real-world scale and complexity, enabling these dynamics to be appropriately represented as multi-scale phenomena. Second, we formulate predictive analysis problems as questions concerning the reachability of diffusion events, and present a novel "altitude function" method for assessing reachability *without simulating system trajectories*. The altitude function technique is both mathematically rigorous and computationally tractable, thereby permitting the derivation of provably-correct assessments for complex, large-scale systems. Third, the S-HDS model and altitude function analytics are used to characterize the importance of *meso-scale* network features, specifically network community and core-periphery structures, for understanding diffusion processes and predicting their fates. This characterization, in turn, forms the foundation for developing a new machine learning-based classification algorithm which employs these network dynamics features for accurate early warning analysis. Additionally, we evaluate the efficacy of this early warning algorithm through three empirical case studies investigating: 1.) the propagation of political "memes" [38] over social media networks, 2.) warning analysis for large-scale mobilization and protest events, and 3.) early warning for politically-motivated cyber attacks. These empirical studies illustrate the effectiveness of the proposed early warning methodology and demonstrate the significant predictive power of meso-scale network metrics for social diffusion processes. Moreover, the results indicate that the proposed algorithm provides a readily-implementable Web-based tool for early warning analysis for important classes of security-relevant diffusion events.

## 2. Early Warning Methodology

This section begins by defining the class of early warning problems of interest, then presents a brief, intuitive summary of the proposed social diffusion modeling and predictive analysis procedure, and finally describes the early warning indicators identified through this analytic proce-dure and the warning algorithm that is derived based on these results. A detailed mathematical presentation of the modeling and analysis methods is provided in Appendices One and Two.

**2.1 Problem Formulation**

The objective of this paper is to develop a scientifically-rigorous, practically-implementable methodology for performing early warning analysis for social diffusion events. Roughly speaking, we suppose that some "triggering event" has taken place or contentious issue is emerging, and we wish to determine, as early as possible, whether this event or issue will ultimately generate a large, self-sustaining reaction, involving the diffusion of discussions and actions through a substantial segment of a population, or will instead quickly dissipate. An illustrative example of the basic idea is provided by the contrasting reactions to 1.) the publication in September 2005 of cartoons depicting Mohammad in the Danish newspaper *Jyllands-Posten*, and 2.) the lecture given by Pope Benedict XVI in September 2006 quoting controversial material concerning Islam. While each event appeared at the outset to have the potential to trigger significant protests, the "Danish cartoons" incident ultimately led to substantial Muslim mobilization, including massive protests and considerable violence, while outrage triggered by the pope lecture quickly subsided with essentially no violence. It would obviously be very useful to have the capability to distinguish these two types of reaction as early in the event lifecycle as possible.

In order to state the early warning problem more precisely, we make a few assumptions:

- We suppose that the triggering event or emerging situation is given. Note that this is often the case in national security settings, and that additionally there exist techniques for *discovering* such events or issues in an automated or semi-automated manner [e.g., 24,27].
- It is assumed that data are available which provide a view of the early reaction of a relevant population to the trigger or issue of interest. These data can be only indirectly related to the event; for example, in this paper the primary data source is social media discussions (e.g., blog posts) while the events of interest are "real-world" activities such as protests.
- It is expected that the "customer" for the analysis provides at least qualitative definitions of the population of interest and the scale of reaction for which a warning is desired. Thus, for instance, in the example above, it might be of interest to anticipate Muslim reaction to the triggering incident, and to obtain a warning alert if the reaction is likely to eventually include self-sustaining, violent protests.

We formulate the early warning problem as a classification task. More specifically, given a triggering incident, one or more information sources which reflect (perhaps indirectly) the reaction to this trigger by a population of interest (e.g., social media discussions, intelligence reporting), and a definition for what constitutes an "alarming" reaction, the goal is to design a classifier which accurately predicts, as early as possible, whether or not reaction to the event will ultimately become alarming. Note that a more mathematically precise statement of this warning problem is given in Appendix Two. Observe that this type of warning analysis is both important in applications and "easier" to accomplish than more standard prediction or forecasting goals. Consider, as a familiar non-security example, the case of movie success. It is shown in [14-16] that it is likely to be impossible to predict movie revenues, even very roughly, based on the intrinsic information available concerning the movie ex ante (e.g., personnel, genre, critic reviews). However, we have demonstrated that it *is* possible to identify early indicators of movie success, such as temporal patterns in pre-release "buzz", and to use these indicators to accurately predict ultimate box office revenues [39]. Recent research indicates that this result holds more generally, so that it may be more scientifically-sensible in many domains to pursue early warning rather than ex ante prediction goals [14-28].

**2.2 S-HDS Social Diffusion Model**

In social diffusion, individuals are affected by what others do. This is easy to visualize in the case of disease transmission, with infections being passed from person to person. Information, innovations, behaviors, and so on can also propagate through a population, as individuals become aware of a new piece of information or an activity and are persuaded of its relevance and utility through their social and information networks. The dynamics of social diffusion can therefore depend upon the topological features of the pertinent networks, such as the presence of highly connected blogs in a social media network (see, e.g., [4]). Indeed, social scientists have developed extensive theories explaining the role of social networks in the dynamics of social diffusion and mobilization (see the books [2-4] and the references therein, and also Appendix One, for discussions of this work). This dependence suggests that, in order to understand the predictability of social diffusion phenomena and in particular to identify features which possess predictive power, it is necessary to conduct the analysis using social and information network models with realistic topologies.

The social diffusion models examined in this study possess networks with three topological properties that are ubiquitous in real-world social and information networks and which have the potential to impact diffusion dynamics [36]:

- *transitivity* – the property that the network neighbors of a given individual have a heightened probability of being connected to one another;
- *community structure* – the presence of densely connected groupings of individuals which have only relatively few links to other groups;
- *core-periphery structure* – the presence of a small group of "core" individuals which are densely connected to each other and are also close to the other individuals in the network.

Additionally, we permit our network models to possess *right-skewed degree distributions,* in which most individuals have only a few network neighbors while a few individuals have a great many neighbors, as such networks are common in online settings. The manner in which the communities and the core-periphery are arranged will be said to define the network's *meso-scale* structure. For convenience of exposition, the subsets of individuals specified by a partitioning of the network into communities and into a core and periphery will sometimes be referred to as the *partition elements*, and the collection of these (community and core-periphery) subsets will be called the *network partition*.

In order to deal effectively with networks possessing realistic topologies, and in particular to represent and analyze the way social dynamics is affected by the meso-scale structure, we model social diffusion in a manner which explicitly separates the individual, or "micro", dynamics from the collective dynamics. More specifically, we adopt a multi-scale modeling framework consisting of three network scales:

- a *micro-scale*, for modeling the behavior of individuals;
- a *meso-scale*, which represents the interaction dynamics of individuals within the same network partition element (community or core/periphery);
- a *macro-scale*, which characterizes the interaction between partition elements.

The micro-scale quantifies the way individuals combine their own inherent preferences or attributes with the influences of others to arrive at their chosen courses of action. It is shown in Appendix One that separating the micro-scale dynamics from the meso- and macro-scale activity permits the dependence of this decision-making process on the social network to be characterized in a surprisingly straightforward way. The meso- and macro-scale components of the proposed modeling framework together quantify the way the decision-making processes of individuals interact to produce collective behavior at the population level. The role of the meso-scale model is to quantify and illuminate the manner in which behaviors *within* each network partition element (communities, core or periphery), while the macro-scale model captures the interactions *between* these elements. The primary assumptions are that interactions between individuals belonging to the same network partition element can be modeled more simply than those between individuals from distinct partition elements, and that the latter interactions are constrained by the "meta-network" which defines the dependencies between the partition elements.

This perspective offers a number of advantages. For example, at the micro-scale it is possible to unify behaviors which appear different phenomenologically but actually possess equivalent dynamics. We show in Appendix One that the social dynamics associated with classical "utility-maximizing" behavior and those arising from individuals attempting to infer information by observing the actions of others can be represented with the *same* micro-scale model. Addi-tionally, separating the individual and collective dynamics supports efficient and flexible model building and simplifies the process of estimating model components from empirical data [39]. Dividing the collective dynamics into meso- and macro-scales also provides a mathematically-tractable, sociologically-sensible means of representing complex social network dynamics. For instance, because network communities are topological structures corresponding to localized social settings in the real world, determined by workplace, family, physical neighborhood, and so on, it is natural both mathematically and sociologically to model the interactions of individuals *within* communities as qualitatively different (e.g., more frequent and homogeneous) than those *between* communities.

Developing a mathematically-rigorous, expressive, scalable, and computationally-tractable framework within which multi-scale social network diffusion models can be constructed is, of course, a challenging undertaking. Recent work in systems biology has demonstrated that stochastic hybrid dynamical systems (S-HDS) provide a useful mathematical formalism with which to represent biological network dynamics that possess multiple temporal and spatial scales [29-33]. An S-HDS is a feedback interconnection of a discrete-state stochastic process, such as a Markov chain, with a family of continuous-state stochastic dynamical systems [34]. Thus the discrete system dynamics depends on the continuous system state, perhaps because different regions of the continuous state space are associated with different matrices of Markov state transition probabilities, and the particular continuous system which is "active" at a given time depends on the discrete system state. Combining discrete and continuous dynamics in this way provides an effective framework for modeling the dynamics of the complex, highly-evolved networks that are ubiquitous in biological systems [35]. For example, the rigorous yet tractable integration of switching behavior with continuous dynamics enabled by the S-HDS model allows accurate and efficient representation of biological phenomena evolving over disparate temporal scales [29-31] and spatial scales [32,33].

Inspired by this work, in this paper we apply the S-HDS framework to social diffusion dynamics evolving over multiple *network* scales. Appendix One provides a detailed discussion of the proposed S-HDS social diffusion model and demonstrates the effectiveness with which this formalism captures multi-scale network dynamics. As an intuitive illustration of the way S-HDS enable complex network phenomena to be efficiently represented, consider the task of modeling diffusion on a network that possesses community structure. As shown in Figure 1, this diffusion consists of two components: 1.) *intra-community dynamics*, involving frequent interactions between individuals within the same community and the resulting gradual change in the concentrations of "infected" (red) individuals, and 2.) *inter-community dynamics*, in which the "infection" jumps from one community to another, for instance because an infected individual "visits" a new community. S-HDS models offer a natural framework for representing these dynamics, with the S-HDS continuous system modeling the intra-community dynamics (e.g., via stochastic differential equations), the discrete system capturing the inter-community dynamics (e.g., using a Markov chain), and the interplay between these dynamics being represented by the S-HDS feedback structure. A detailed description of the manner in which S-HDS models can be used to capture social diffusion on networks with realistic topologies is given in Appendix One.

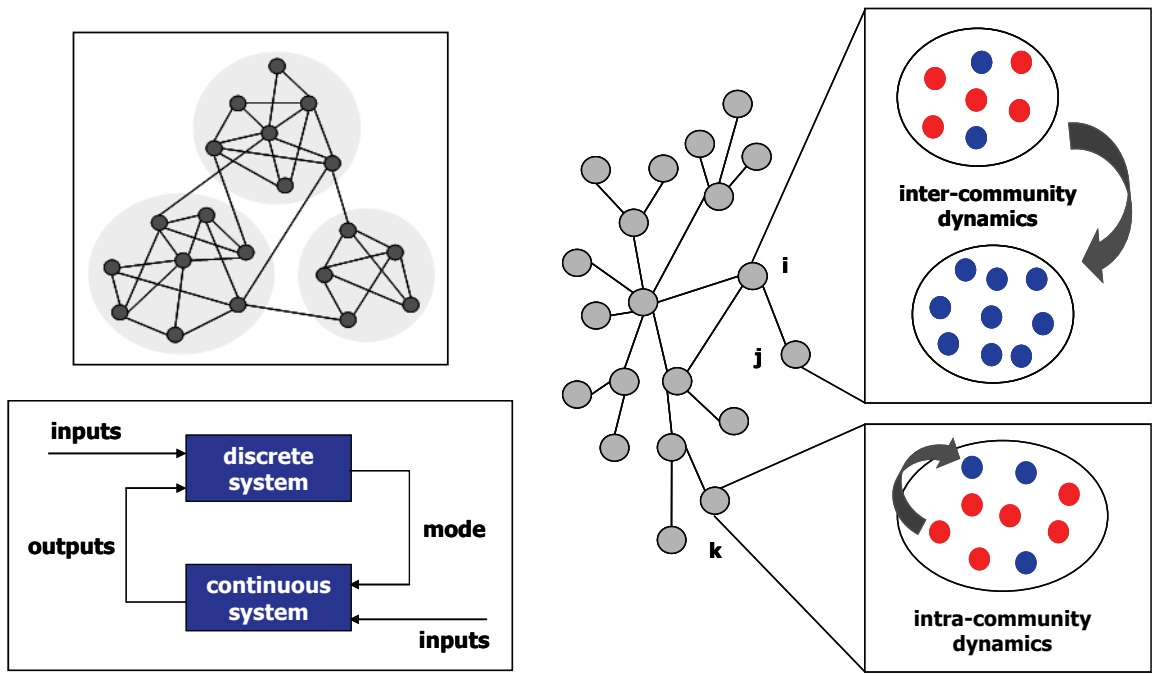

**Figure 1.** Modeling diffusion on networks with community structure via S-HDS. The cartoon at top left depicts a network with three communities. The cartoon at right illustrates diffusion *within* a community k and *between* communities i and j. The schematic at bottom left shows the basic S-HDS feedback structure; the discrete and continuous systems in this framework model the inter-community and intra-community diffusion dynamics, respectively.

### 2.3 Predictability Assessment

One hallmark of social diffusion processes is their ostensible unpredictability: phenomena from hits and flops in cultural markets to financial system bubbles and crashes to political upheavals appear resistant to predictive analysis (although there is no shortage of ex post explanations for their occurrence!). It is not difficult to gain an intuitive understanding of the basis for this unpredictability. Individual preferences and susceptibilities are mapped to collective outcomes through an intricate, dynamical process in which people react individually to an environment consisting largely of others who are reacting likewise. Because of this feedback dynamics, the collective outcome can be quite different from one implied by a simple aggregation of indi-vidual preferences; standard prediction methods, which typically are based on such aggregation ideas, do not capture these dynamics and therefore are often unsuccessful.

This section provides a brief, intuitive introduction to a systematic approach to assessing the predictability of social diffusion processes and identifying process observables which have exploitable predictive power (see Appendix Two, and also [17,39], for the mathematical details). Consider a simple model for product adoption, in which individuals combine their own preferences and opinions regarding the available options with their observations of the actions of others to arrive at their decisions about which product to adopt. As discussed above, it can be quite difficult to determine which characteristics of the process by which adoption decisions propagate, if any, are predictive of things like the speed or ultimate reach of the propagation [15-17]. In Appendix Two we propose a mathematically rigorous approach to predictability assessment which, among other things, permits identification of features of social dynamics which should have predictive power. We now summarize this assessment methodology.

The basic idea behind the proposed approach to predictability analysis is simple and natural: we assess predictability by answering questions about the reachability of diffusion events. To obtain a mathematical formulation of this strategy, the behavior about which predictions are to be made is used to define the system *state space subsets of interest* (SSI), while the particular set of candidate measurables under consideration allows identification of the *candidate starting set* (CSS), that is, the set of states and system parameter values which represent initializations that are consistent with, and equivalent under, the presumed observational capability. As a simple example, consider an online market with two products, A and B, and suppose the system state variables consist of the current market share for A, ms(A), and the rate of change of this market share, r(A) (ms(B) and r(B) are not independent state variables because ms(A) + ms(B) = 1 and r(A) + r(B) = 0); let the parameters be the advertising budgets for the products, bud(A) and bud(B). The producer of item A might find it useful to define the SSI to reflect market share dominance by A, that is, the subset of the two-dimensional state space where ms(A) exceeds a specified threshold (and r(A) can take any value). If only market share and advertising budgets can be measured then the CSS is the one-dimensional subset of state-parameter space consisting of the initial magnitudes for ms(A), bud(A), and bud(B), with r(A) unspecified (the one-dimensional "uncertainty" in the CSS reflects the fact that r(A) is not measurable).

Roughly speaking, the proposed approach to predictability assessment involves determining how probable it is to reach the SSI from a CSS and deciding if these reachability properties are compatible with the prediction goals. If a system's reachability characteristics are incompatible with the given prediction question – if, say, "hit" and "flop" states in the online market example are both fairly likely to be reached from the CSS – then the situation is deemed unpredictable. This setup permits the identification of candidate predictive measurables: these are the measur-able states and/or parameters for which predictability is most sensitive (see Appendix Two). Continuing with the online market example, if trajectories with positive early market share rates r(A) are much more likely to yield market share dominance for A than are trajectories with negative early r(A), then the situation is unpredictable (because the outcome depends sensitively on r(A) and this quantity is not measured). Moreover, this analysis suggests that market share rate is likely to possess predictive power, so it may be possible to increase predictability by adding the capacity to measure this quantity.

A key element of this approach to predictability assessment is the proposed method of estimating the probability of reaching the SSI from a CSS. Note that in a typical assessment such estimates must be computed for several CSS in order to adequately explore the space of candidate predictive features, so that it is crucial to perform these estimates efficiently. In Appendix Two we develop an "altitude function" approach to this reachability problem, in which we seek a scalar function of the system state that permits conclusions to be made regarding reachability *without computing system trajectories*. We refer to these as altitude functions to provide an intuitive sense of their analytic role: if some measure of "altitude" is low on the CSS and high on an SSI, and if the expected rate of change of altitude along system trajectories is nonincreasing, then it is unlikely for trajectories to reach this SSI from the CSS. Moreover, the difference in altitudes between the CSS and SSI gives a measure of the probability of reaching the latter from the former. Because the reach probability is computed for *sets* of states without simulating system trajectories, the altitude function method offers an extremely efficient way to explore the space of candidate predictive features.

We have applied the predictability assessment methodology summarized above to the social diffusion prediction problem, and we now summarize the main conclusions of this study; a more complete discussion of this investigation is given in Appendix Two. The analysis uses the mathematically rigorous predictability assessment procedure summarized above, in combination with empirically-grounded S-HDS models for social dynamics, to characterize the predictability of social diffusion on networks with realistic degree distributions, transitivity, community structure, and core-periphery structure. The main finding of the study, from the perspective of the present paper, is that the predictability of these diffusion models depends crucially upon social and information network topology, and in particular on the community and core-periphery structures of these networks.

In order to describe these theoretical results more quantitatively and leverage them for prediction, it is necessary to specify mathematical definitions for network communities and core-periphery structure. There exist several qualitative and quantitative definitions for the concept of community structure in networks. Here we adopt the *modularity-based* definition proposed in [40], whereby a good partitioning of a network's vertices into communities is one for which the number of edges between putative communities is smaller than would be expected in a random partitioning. To be concrete, a modularity-based partitioning of a network into two communities maximizes the modularity Q, defined as

$$Q = s^T B s / 4m,$$

where m is the total number of edges in the network, the partition is specified with the elements of vector s by setting $s_i = 1$ if vertex i belongs to community 1 and $s_i = -1$ if it belongs to community 2, and the matrix B has elements $B_{ij} = A_{ij} - k_i k_j / 2m$, with $A_{ij}$ and $k_i$ denoting the network adjacency matrix and degree of vertex i, respectively. Partitions of the network into more than two communities can be constructed recursively [40]. Note that modularity-based community partitions can be efficiently computed for large social networks, and can be constructed even with incomplete network topology data [39].

With this definition in hand, we are in a position to present the first candidate predictive feature nominated by the theoretical predictability assessment: the presence of early diffusion activity in numerous distinct network communities should be a reliable predictor that the ultimate reach of the diffusion will be large (see Appendix Two). In what follows, propagation dynamics which possess this characteristic will be said to exhibit *significant early dispersion across network communities.* Note that this measure should be more predictive than the early volume of diffusion activity (the latter has recently become a fairly standard measure [e.g., 19,20]). A cartoon illustrating the basic idea behind this result is given in Figure 2.

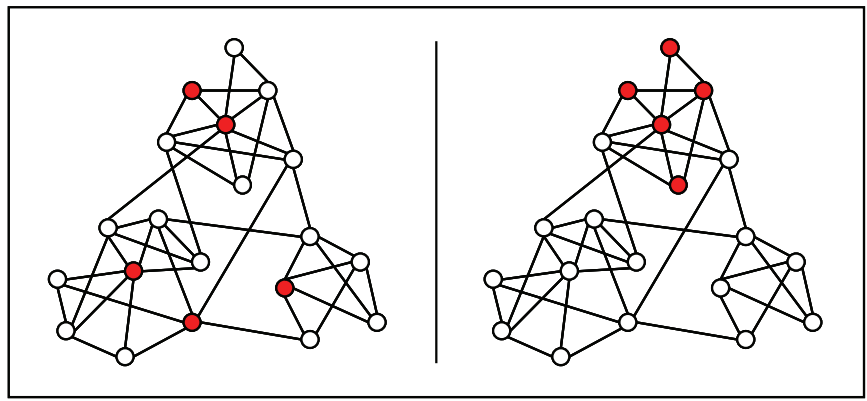

**Figure 2.** Early dispersion across communities is predictive. The cartoon illustrates the predictive feature associated with community structure: social diffusion initiated with five "seed" individuals is much more likely to propagate widely if these seeds are dispersed across three communities (left) rather than concentrated within a single community (right). Note that in Appendix Two this result is established for networks of realistic scale and not simply for "toy" networks like the one shown here.

Analogously to the situation with network communities, there exists a wide range of qualitative and quantitative descriptions of the core-periphery structure found in real-world networks. Here we adopt the characterization of network core-periphery which results from *k-shell decomposition*, a well-established technique in graph theory that is summarized in, for instance, [41]. To partition a network into its k-shells, one first removes all vertices with degree one, repeating this step if necessary until all remaining vertices have degree two or higher; the removed vertices constitute the 1-shell. Continuing in the same way, all vertices with degree two (or less) are recursively removed, creating the 2-shell. This process is repeated until all vertices have been assigned to a k-shell. The shell with the highest index, the $k_{max}$-shell, is deemed to be the core of the network.

Given this definition, we are in a position to report the second candidate predictive feature nominated by our theoretical predictability assessment: early diffusion activity within the network $k_{max}$-shell should be a reliable predictor that the ultimate reach of the diffusion will be significant (see Appendix Two). In particular, this measure should be more predictive than the early volume of diffusion activity. An intuitive illustration of this result is depicted in Figure 3.

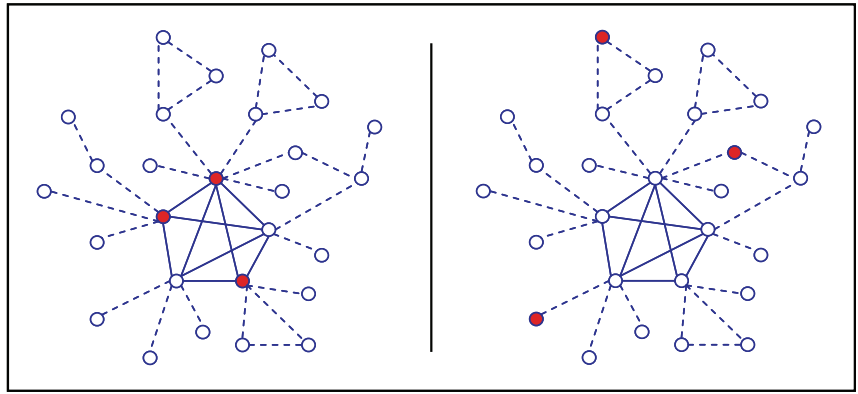

**Figure 3.** Early diffusion within the core is predictive. The cartoon illustrates the predictive feature associated with k-shell structure: social diffusion initiated with three "seed" individuals is much more likely to propagate widely if these seeds reside within the network's core (left) rather than at its periphery (right). Note that in Appendix Two this result is established for networks of realistic scale and not simply for "toy" networks like the one shown here.

**2.4 Early Warning Method**

We are now in a position to present an early warning method which is capable of accurately predicting, very early in the lifecycle of a diffusion process of interest, whether or not the process will propagate widely. We adopt a machine learning-based classification approach to this prob-lem: given a triggering incident, one or more information sources which reflect the reaction to this trigger by a population of interest, and a definition for what constitutes an "alarming" reaction, the goal is to learn classifier that accurately predicts, as early as possible, whether or not reaction to the event will ultimately become alarming. The classifier used in the empirical studies described in this paper is the Avatar ensembles of decision trees (A-EDT) algorithm [42]. Other classification algorithm were also explored to allow the robustness of the proposed early warning approach to be evaluated, and these alternative methods produced qualitatively similar results [39]. Prediction accuracy in all tests is estimated using standard N-fold cross-validation, in which the set of diffusion events of interest is randomly partitioned into N subsets of equal size, and the A-EDT algorithm is successively "trained" on N−1 of the subsets and "tested" on the held-out subset in such a way that each of the N subsets is used as the test set exactly once.

A key aspect of the proposed approach to early warning analysis is determining which characteristics of the social diffusion event of interest, if any, possess exploitable predictive power. We consider three classes of features:

- *intrinsics-based features* – measures of the inherent properties and attributes of the "object" being diffused;
- *simple dynamics-based features* – metrics which capturing simple properties of the diffusion dynamics, such as the early extent of the diffusion and the rate at which the diffusion is propagating;
- *network dynamics-based features* – measures that characterize the way the early diffusion is progressing relative to topological properties of the underlying social and information networks (e.g., community structure).

Consider, as an illustrative example, the diffusion of "memes", that is, short textual phrases which propagate relatively unchanged online (e.g., 'lipstick on a pig'). Suppose it is of interest to predict which memes will "go viral", appearing in thousands of blog posts, and which will not. In this case, intrinsic-based features could include language measures, such as the sentiment or emotion expressed in the text surrounding the memes in blog posts or news articles. Simple dynamics-based features for memes might measure the cumulative number of posts or articles mentioning the meme of interest at some early time $\tau$ and the rate at which this volume is increasing. Network dynamics-based features might count the cumulative number of network communities in a blog graph $G_B$ that contain at least one post which mentions the meme by time $\tau$ and the number of blogs in the $k_{max}$-shell of $G_B$ that, by time $\tau$, contain at least one post mentioning the meme. Alternatively, in the case of an epidemic, the intrinsic-based features could include the infectivity of the pathogen, simple dynamics-based features might capture the number of individuals infected by the disease in the early stages of the outbreak, and network dynamics-based features could include metrics that characterize the way the epidemic is progressing over the communities of relevant social and transportation networks.

The proposed approach to early warning analysis is to collect features from these classes for the event of interest, input the feature values to the (trained) A-EDT classifier, and then run the classifier to generate the warning prediction (i.e., a forecast that the event is expected to become 'alarming' or remain 'not alarming'). In the algorithm presented below this procedure in specified in general terms; more specific instantiations of the procedure are presented in the discussions of the three case studies in Section 3. In what follows it is assumed that the primary source of information concerning the event of interest is social media, as that is emerging as a very useful data source for predictive analysis [e.g., 17-24,26,27]. However, the analytic process is quite similar when other data sources (e.g., intelligence reporting) are employed [24].

Thus we have the following early warning algorithm:

**Algorithm EW**

Given: a triggering incident, a definition for what constitutes an 'alarming' reaction, and a set of social media sites (e.g., blogs) B which are relevant to early warning task.

Initialization: train the A-EDT classifier on a set of events which are qualitatively similar to the triggering event of interest and are labeled as 'alarming' or 'not alarming' according to the definition given above (see the case study discussions for additional details on this training process).

Procedure:

1. Assemble a lexicon of keywords L that pertain to the triggering event under study.
2. Conduct a sequence of blog graph crawls and construct a time series of blog graphs $G_B(t)$. For the lexicon L and each time period t, label each blog in $G_B(t)$ as 'active' if it contains a post mentioning any of the keywords in L and 'inactive' otherwise.
3. Form the union $G_B = \cup_t G_B(t)$, partition $G_B$ into network communities and into k-shells, and map the partition element structure of $G_B$ back to each of the graphs $G_B(t)$.
4. Compute the values of appropriate measures for the intrinsics, simple dynamics, and network dynamics features for each of the graphs $G_B(t)$.
5. Apply the A-EDT classifier to the available time series of features, that is, the features obtained from the sequence of blog graphs $\{G_B(t_0), …, G_B(t_p)\}$, where $t_0$ and $t_p$ are the triggering event time and present time, respectively. Issue an early warning alert if the classifier output is 'alarming'.

We now offer additional details concerning this procedure; more application-specific discussions of the methodology are provided in the case studies in Section 3. Identifying appropriate keywords in Step 1 can be accomplished with the help of subject matter experts and also through various automated means (e.g., via meme analysis [38,27]). Step 2 is by now standard, and various tools exist which can perform these tasks [e.g., 43]. In Step 3, blog network communities are identified with a modularity-based community extraction algorithm applied to the blog graph [40], while the decomposition of the graph into its k-shells is achieved through standard methods [41]. The particular choices of metrics for the intrinsics, simple dynamics, and network dynamics features computed in Step 4 tend to be problem specific, and typical examples are given in the case studies below. It is worth noting, however, that we have found it useful in a range of applications to quantify the dispersion of activity over the communities of $G_B(t)$ using a blog entropy measure BE:

$$BE(t) = -\Sigma_i\, f_i(t)\, \log(f_i(t)),$$

where $f_i(t)$ is the fraction of total posts containing one or more keywords and made during interval t which occur in community i. Finally, in Step 5 the feature values obtained in Step 4 serve as inputs to the A-EDT classifier and the output is used to decide whether an alert should be issued.

## 3. Case Studies

This section applies Algorithm EW to three early warning case studies involving social phenomena that have proved to be both practically important and challenging to analyze: 1.) diffusion of information through social media, 2.) mobilization/protest events response to "triggering" incidents, and 3.) planning/coordination/execution of politically-motivated cyber attacks.

### 3.1 Case Study One: Meme Diffusion

The goal of this case study is to apply Algorithm EW to the task of predicting whether or not a given "meme", that is, a short textual phrase which propagates relatively unchanged online, will "go viral". Our main source of data on meme dynamics is the publicly available datasets archived at http://memetracker.org [44] by the authors of [38]. Briefly, the archive [44] contains time series data characterizing the diffusion of ~70 000 memes through social media and other online sites during the five month period between 1 August and 31 December 2008. We are interested in using Algorithm EW to distinguish successful and unsuccessful memes early in their lifecycle. More precisely, the task of interest is to classify memes into two groups – those which will ultimately be successful (acquire more than S posts) and those that will be unsuccessful (attract fewer than U posts) – very early in the meme lifecycle.

To support an empirical evaluation of the utility of Algorithm EW for this problems, we downloaded from [44] the time series data for slightly more than 70 000 memes. These data contain, for each meme M, a sequence of pairs $(t_1, URL_1)_M, (t_2, URL_2)_M, \ldots, (t_T, URL_T)_M$, where $t_k$ is the time of appearance of the kth blog post or news article that contains at least one mention of meme M, URL$_k$ is the URL of the blog or news site on which that post/article was published, and T is the total number of posts that mention meme M. From this set of time series we randomly selected 100 "successful" meme trajectories, defined as those corresponding to memes which attracted at least 1000 posts during their lifetimes, and 100 "unsuccessful" meme trajectories, defined as those whose memes acquired no more than 100 total posts. It is worth noting that, in assembling the data in [44], all memes which received fewer than 15 total posts were deleted, and that ~50% of the remaining memes have <50 posts; thus the large majority of memes are unsuccessful by our definition (as well as according to the criteria of most applications [38,27]).

Two other forms of data were collected for this study: 1.) a large Web graph which includes websites (URLs) that appear in the meme time series, and 2.) samples of the text surrounding the memes in the posts which contain them. More specifically, we sampled the URLs appearing in the time series for our set of 200 successful and unsuccessful memes and performed a Web crawl that employed these URLs as "seeds". This procedure generated a Web graph, denoted $G_B$, that consists of approximately 550 000 vertices/websites and 1.4 million edges/hyperlinks, and includes essentially all of the websites which appear in the meme time series. To obtain samples of text surrounding memes in posts, we randomly selected ten posts for each meme and then extracted from each post the paragraph which contains the first mention of the meme.

Recall that Algorithm EW employs three types of features: intrinsics-based, simple dynamics-based, and network dynamics-based. We now describe the instantiation of each of these feature classes for the meme problem. Consider first the intrinsics-based features, which for the meme application become language-based measures. Each "document" of text surrounding a meme in its (sample) posts is represented by a simple "bag of words" feature vector $x \in \Re^{|V|}$, where the entries of x are the frequencies with which the words in the vocabulary set V appear in the document. A very simple way to quantify the sentiment or emotion of a document is through the use of appropriate lexicons. Let $s \in \Re^{|V|}$ denote a lexicon vector, in which each entry of s is a numerical "score" quantifying the sentiment/emotion intensity of the corresponding word in the vocabulary V. The aggregate sentiment/emotion score of document x can be computed as

$$\text{score}(x) = s^T x / s^T 1,$$

where 1 is a vector of ones. Thus score(.) estimates the sentiment or emotion of a document as a weighted average of the sentiment or emotion scores for the words comprising the document. (Note that if no sentiment or emotion information is available for a particular word in V then the corresponding entry of s is set to zero.)

To characterize the emotion content of a document we use the Affective Norms for English Words (ANEW) lexicon, which consists of 1034 words that were assigned numerical scores with respect to three emotional "axes" – happiness, arousal, and dominance – by human subjects [45].

Previous work had identified this set of words to bear meaningful emotional content [45]. Positive or negative sentiment is quantified by employing the "IBM lexicon", a collection of 2968 words that were assigned {positive, negative} sentiment labels by human subjects [46]. This simple approach generates four language features for each meme: the happiness, arousal, dominance, and positive/negative sentiment of the text surrounding that meme in the (sample) posts containing it. As a preliminary test, we computed the mean emotion and sentiment of content surrounding the 100 successful and 100 unsuccessful memes in our dataset. On average the text surrounding successful memes is happier, more active, more dominant, and more positive than that surrounding unsuccessful memes, and this difference is statistically significant ($p<0.0001$). Thus it is at least plausible that these four language features may possess some predictive power regarding meme success.

Consider next two simple dynamics-based features, defined to capture the basic characteristics of the early evolution of meme post volume:

- #posts($\tau$) – the cumulative number of posts mentioning the given meme by time $\tau$ (where $\tau$ is small relative to the typical lifespan of memes);
- post rate($\tau$) – a simple estimate of the rate of accumulation of such posts at time $\tau$.

Here we adopt a simple finite difference definition for post rate given by post rate($\tau$) = (#posts($\tau$) – #posts($\tau/2$)) / ($\tau/2$); of course, more robust rate estimates could be used.

The simple dynamics-based measures of early meme diffusion defined above, while potentially useful, do not characterize the manner in which a meme propagates over the underlying social or information networks. Recall that the predictability assessment summarized in Section 2.3 suggests that both early dispersion of diffusion activity across network communities and early diffusion activity within the network core ought to be predictive of meme success. The insights offered by this theoretical analysis motivate the definition of two network dynamics-based features for meme prediction:

- community dispersion($\tau$) – the cumulative number of network communities in the blog graph $G_B$ that, by time $\tau$, contain at least one post which mentions the meme;
- #k-core blogs($\tau$) – the cumulative number of blogs in the $k_{max}$-shell of blog graph $G_B$ that, by time $\tau$, contain at least one post which mentions the meme.

These quantities can be efficiently computed using fast algorithms for partitioning a graph into its communities and for identifying a graph's $k_{max}$-shell [39]. Thus these features are readily computable even for very large graphs.

We now summarize the results of this case study. First, using only the four language features with the A-EDT classifier to predict which memes will be successful yields a prediction accuracy of 66.5% (ten-fold cross-validation). Since simply guessing "successful" for all memes gives an accuracy of 50%, it can be seen that these simple language intrinsics are not very predictive. For completeness it is mentioned that the ANEW score for "arousal" and the IBM measure of sentiment are the most predictive of these four features. In contrast, the features characterizing the early network dynamics of memes possess significant predictive power, and in fact are useful even if only very limited early time series is available for use in prediction. More quantitatively, applying Algorithm EW with the four meme dynamics features produces the following results (ten-fold cross-validation):

- $\tau = 12$hr, accuracy = 84%, most predictive features: 1.) community dispersion, 2.) #k-core blogs, 3.) #posts;
- $\tau = 24$hr, accuracy = 92%, most predictive features: 1.) community dispersion, 2.) post rate, 3.) #posts;
- $\tau = 48$hr, accuracy = 94%, most predictive features: 1.) community dispersion, 2.) post rate, 3.) #posts.

These results show that useful predictions can be obtained *within the first twelve hours* after a meme is detected (this corresponds to 0.5% of the average meme lifespan), and that accurate prediction is possible after about a day or two. Note also that, as has been found with other social dynamics phenomena [e.g., 16-18], dynamics features appear to be more predictive than "intrinsics", at least for the features employed here.

It is worth mentioning that the fact that a particular meme goes viral does not imply that it will influence behavior in the real world. The next two case studies focus on the important issue of behavioral consequences of information diffusion.

### 3.2 Case Study Two: Mobilization and Protest

There is considerable interest to develop methods for distinguishing successful mobilization and protest events, that is, mobilizations that become large and self-sustaining, from unsuccessful ones early in their lifecycle. It is natural to pose this question as an early warning problem and to approach it using Algorithm EW. In order to examine the efficacy of this approach, we collected together fourteen recent events, each of which appeared at the outset to have the potential to trigger significant protests. This set of events contains seven triggering incidents which ultimately led to substantial mobilization, including massive protests and significant violence, and seven triggers with reactions that subsided quickly with essentially no violence. Taken together, these events provide a useful setting for testing the applicability of Algorithm EW to mobilization/protest phenomena.

The events employed in this study are listed below.

Triggers leading to significant mobilization/protest:

- Quran desecration, May 2005;
- first Danish cartoons, September 2005 to March 2006;
- Egypt DVD release, October 2005;
- France riots, October and November 2005;
- anti-Ahmadiyya protests, June and July 2008;
- U.S Republican National Convention, September 2008;
- Israel/Gaza event, December 2008 to January 2009.

Triggers not leading to significant mobilization/protest:

- Abu Ghraib news release, April and May 2004;
- Pope lecture, September 2006;
- Salman Rushdie knighting, June 2007;
- second Danish cartoons, February 2008;
- U.S. Democratic National Convention, August 2008,
- Bali bombers execution, November 2008;
- Jakarta bombings/NM Top blog post, July 2009.

This list is intended merely to identify the fourteen events under study; additional information concerning each incident is given in [39] and the references therein.

As a preliminary examination of the possibility to obtain useful early warning indicators from analysis of social media discussions of these events, we performed Steps 1-4 of Algorithm EW and then plotted the time series for two quantities: 1.) the volume of blog posts mentioning keywords relevant to the events (these keywords were obtained through a simple news search [39]), and 2.) the blog entropy measure $BE(t) = -\Sigma_i\, f_i(t)\, \log(f_i(t))$ associated with the way online mentions of the keywords diffused over the blog graph. Illustrative time series plots are shown in Figure 4. Observe that in the case of the first Danish cartoons event (plot at right) the BE of relevant discussions (blue curve) experiences a dramatic increase a few weeks before the corresponding increase in volume of blog discussions (red curve); this latter increase, in turn, takes place before any violence. In contrast, in the case of the pope event (plot at left), BE of blog discussions is small relative to the cartoons event, and any increase in this measure lags discussion volume. Similar time series plots are obtained for the other twelve events, suggesting that network dynamics-based features, such as dispersion of discussions across blog network communities, may be a useful early indicator for large mobilization events.

To examine this possibility more carefully, we applied Algorithm EW to the task of distinguishing triggers which led to large protests from those that did not. For simplicity, in this case study we did not use any intrinsics-based features (e.g., language metrics) in the A-EDT classi-fier, and instead relied upon the four dynamics-based features defined in Case Study One. In the case of the seven triggering events which led to protest behavior, the blog data made available to Algorithm EW was limited to posts made during the eight week period which ended two weeks before the protests began. For the seven triggers which did not lead to protests, the blog data included all posts collected during the eight week period immediately following the triggering event.

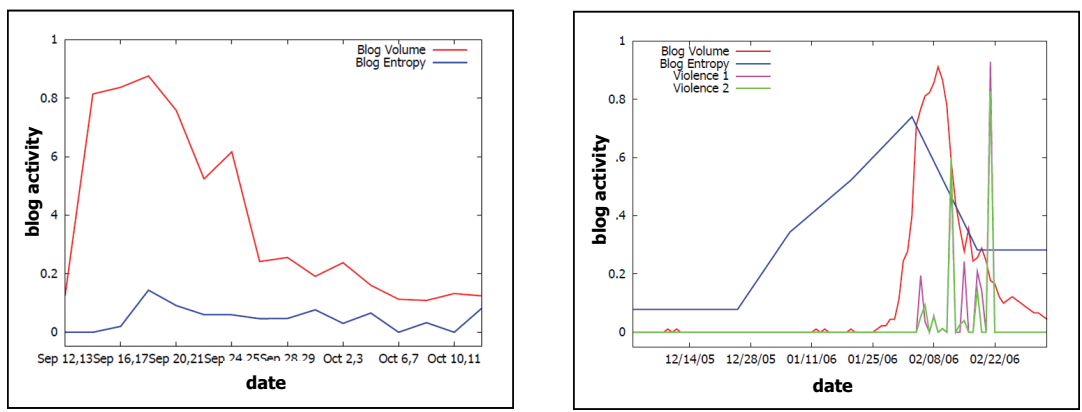

**Figure 4.** Sample results for mobilization/protest case study. The illustrative time series plots shown correspond to the pope event (left) and first Danish cartoons event (right). In each plot, the red curve is blog volume and the blue curve is blog entropy; the Danish cartoon plot also shows two measures of violence (cyan and magenta curves). Note that while the volume and violence data are scaled to allow multiple data sets to be graphed on each plot, the scale for entropy is consistent across plots to enable cross-event comparison.

Because the set of events in this case study included only fourteen incidents, we applied Algorithm EW with two-fold cross-validation. More specifically, the set of incidents was randomly partitioned into two equal subsets, the algorithm was trained on one subset of seven incidents and tested on the other subset, and then the roles of the two data sets were switched. In this evaluation Algorithm EW achieved *perfect* accuracy, correctly distinguishing the 'protest' and 'non-protest' triggers. An examination of the predictive power of the four features used as inputs to the A-EDT classifier reveals that, as suggested by Figure 4, the community dispersion feature was the most predictive measure.

### 3.2 Case Study Three: Cyber Attack Early Warning

This case study explores the ability of Algorithm EW to provide reliable early warning for politically-motivated distributed denial-of-service (DDoS) attacks. Toward this end, we first identified a set of Internet "disturbances" that included examples from three distinct classes of events:

1. successful politically-motivated DDoS attacks – these are the events for which Algorithm EW is intended to give warning with sufficient lead time to allow mitigating actions to be taken;
2. natural events which disrupt Internet service – these are disturbances, such as earthquakes and electric power outages, that impact the Internet but for which it is known that no early warning signal exists in social media;
3. quiet periods – these are periods during which there is social media "chatter" concerning impending DDoS attacks but ultimately no (successful) attacks occurred.

Including in the case study events selected from these three classes is intended to afford a fairly comprehensive test of Algorithm EW. For instance, these classes correspond to 1.) the domain of interest (DDoS attacks), 2.) a set of disruptions which impact the Internet but have no social media warning signal, and 3.) a set of "non-events" which do not impact the Internet but do possess putative social media warning signals (online discussion of DDoS attacks).

We selected twenty events from these three classes:

Politically-motivated DDoS attacks:
- Estonia event in April 2007;
- CNN/China incident in April 2008;
- Israel/Palestine conflict event in January 2009;
- DDoS associated with Iranian elections in June 2009;
- WikiLeaks event in November 2010;
- Anonymous v. PayPal, etc. attack in December 2010;
- Anonymous v. HBGary attack in February 2011.

Natural disturbances:
- European power outage in November 2006;
- Taiwan earthquake in December 2006;
- Hurricane Ike in September 2008;
- Mediterranean cable cut in January 2009;
- Taiwan earthquake in March 2010;
- Japan earthquake in March 2011.

Quiet periods:

Seven periods, from March 2005 through March 2011, during which there were discussions in social media of DDoS attacks on various U.S. government agencies but no (successful) attacks occurred.

For brevity a detailed discussion of these twenty events is not given here; the interested reader is referred to [39] and the references therein for additional information on these disruptions.

We collected two forms of data for each of the twenty events: *cyber data* and *social data*. The cyber data consist of time series of routing updates which were issued by Internet routers during a one month period surrounding each event. More precisely, these data are the Border Gateway Protocol (BGP) routing updates exchanged between gateway hosts in the Autonomous System network of the Internet. The data was downloaded from the publicly-accessible RIPE collection site [47] using the process described in [48] (see [48] for additional details and background information on BGP routing dynamics). The temporal evolution of the volume of BGP routing updates (e.g., withdrawal messages) gives a coarse-grained measure of the timing and magnitude of large Internet disruptions and thus offers a simple and objective way to characterize the impact of each of the events in our collection. The social data consist of time series of social media mentions of cyber attack-related keywords and memes detected during a one month period surrounding each of the twenty events. These data were collected using the procedure specified in Algorithm EW.

As in the preceding case study, we performed a preliminary examination of the possibility to obtain useful early warning indicators from analysis of social media discussions by completing Steps 1-4 of Algorithm EW and plotting the time series for two quantities: 1.) the volume of blog posts mentioning keywords relevant to the events (these keywords were obtained through a simple news search [39]), and 2.) the blog entropy measure $BE(t) = -\Sigma_i f_i(t) \log(f_i(t))$ associated with the way online mentions of the keywords diffused over the blog graph. Illustrative time series plots corresponding to two events in the case study, the WikiLeaks DDoS attack in November 2010 and Japan earthquake in March 2011, are shown in Figure 5. Observe that the time series of BGP routing updates are similar for the two events, with each experiencing a large "spike" at the time of the event. The time series of blog post volume are also similar across the two events, with each showing modest volume prior to the event and displaying a large spike in activity at event time. However, the time series for blog entropy are quite distinct for the two events. Specifically, in the case of the WikiLeaks DDoS the blog entropy (blue curve in Figure 5) experiences a dramatic increase several days before the event, while in the case of the Japan earthquake blog entropy is small for the entire collection period. Similar social media behavior is observed for all events in the case study, suggesting that network dynamics-based features, such as dispersion of discussions across blog network communities, may be a useful early indicator for large mobilization events.

To examine this possibility more carefully, we applied Algorithm EW to the task of distinguishing the seven DDoS attacks from the thirteen other events in the set. For simplicity, in this case study we did not use any intrinsics-based features (e.g., language metrics) in the A-EDT classifier, and instead relied upon the four dynamics-based features defined in Case Study One.

Because the set of events in this case study included only twenty incidents, we applied Algorithm EW with two-fold cross-validation, exactly as described in Case Study Two. In the case of DDoS events, the blog data made available to Algorithm EW was limited to posts made during the five week period which ended one week before the attack. For the six natural disturbances, the blog data included all posts collected during the six week period immediately prior to the event, while in the case of the seven non-events, the blog data included the posts collected during a six week interval which spanned discussions of DDoS attacks on U.S. government agencies.

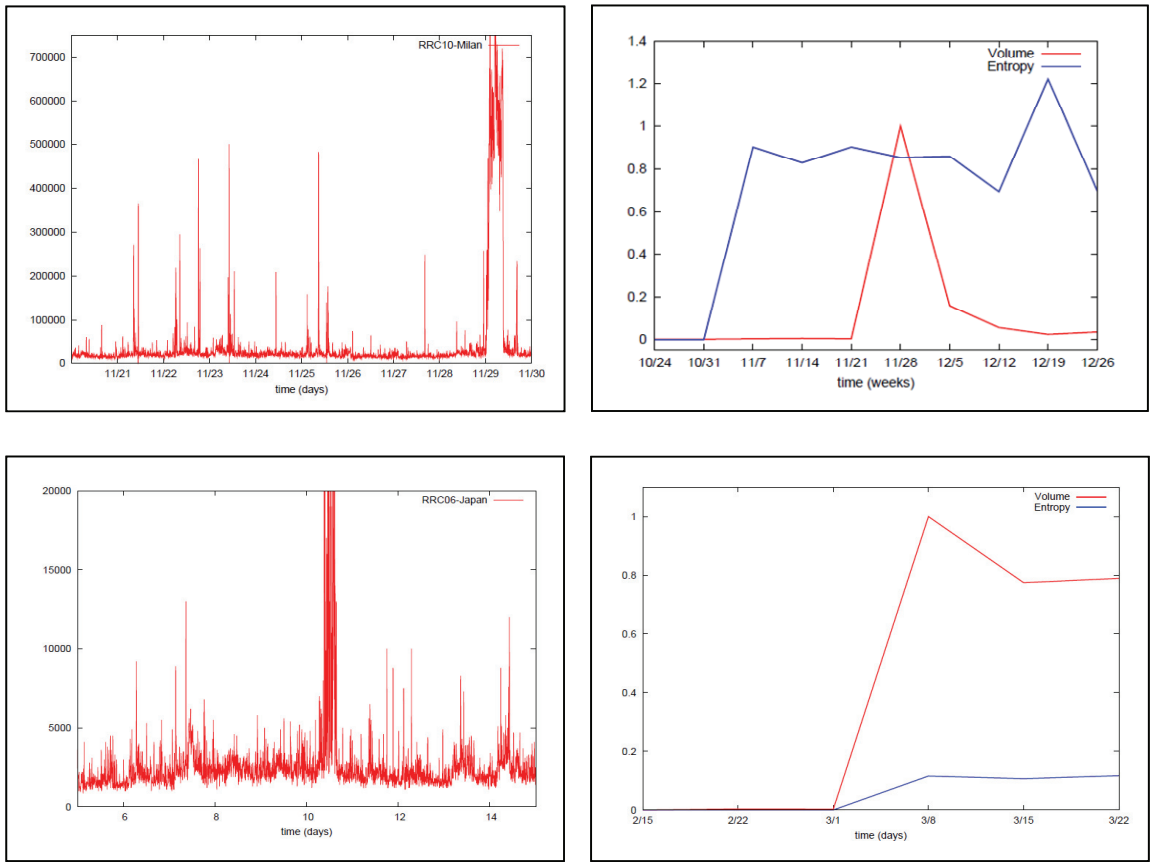

**Figure 5.** Sample results for the DDoS early warning case study. The illustrative time series plots shown correspond to the WikiLeaks event in November 2010 (top row) and the Japan earthquake in March 2011 (bottom row). For each event, the plot at left is the time series of BGP routing updates (note the large increase in updates triggered by the event). The plot at the right of each row is the time series of the social media data, with the red curve showing blog post volume and the blue curve depicting blog entropy. Note that while post volume is scaled for convenient visualization, the scale for entropy is consistent across plots to allow cross-event comparison.

In this evaluation, Algorithm EW achieved *perfect* accuracy, correctly distinguishing the 'attack' and 'non-attack' events. If the test is made more difficult, so that the blog data made available to Algorithm EW for attack events is limited to a four week period that ends two weeks before the attack, the proposed approach still achieves 95% accuracy, An examination of the predictive power of the four features used as inputs to the A-EDT classifier reveals that, as suggested by Figure 5, the community dispersion feature was the most predictive measure. It is worth emphasizing that, in this case study, accurately distinguishing 'attack' from 'non-attack' events is equivalent to providing practically-useful early warning for attack events, because the data which serves as input to Algorithm EW reflects online discussions that took place *prior to* the events under investigation.

## 4. Conclusions

This paper presents a new approach to early warning analysis for social diffusion events. We begin by introducing a biologically-inspired S-HDS model for social dynamics on multi-scale networks, and then perform stochastic reachability analysis with this model to show that the outcomes of social diffusion processes may depend crucially upon the way the early dynamics of the process interacts with the underlying network's meso-scale topological structures. This theoretical finding provides the foundations for developing a machine learning algorithm that enables accurate early warning analysis for diffusion events. The utility of the warning algorithm, and the power of network-based predictive metrics, are demonstrated through empirical case studies involving meme propagation, large-scale protests events, and politically-motivated cyber attacks.

## 5. Acknowledgements

This research was supported by the U.S. Department of Defense, the U.S. Department of Homeland Security, The Boeing Company, and the Laboratory Directed Research and Development program at Sandia National Laboratories. Fruitful discussions regarding aspects of this work with Curtis Johnson of Sandia National Laboratories, Paul Ormerod of Volterra Partners, and Anne Kao of Boeing are gratefully acknowledged.

## A1. Appendix One: S-HDS Social Diffusion Model

In this Appendix we propose a multi-scale structure for modeling social network dynamics, establish a few facts concerning this representation, and introduce an S-HDS formulation of the model that is well-suited for predictive analysis.

### A1.1 Multi-Scale Social Dynamics Model

In many social situations, people are influenced by the behavior of others, for instance because they seek to obtain the benefits of coordinated actions, infer otherwise inaccessible information, or manage complexity in decision-making. Processes in which observing a certain behavior increases an individual's probability of adopting that behavior are often referred to as *positive externality processes* (PEP), and we use that term here. PEP have been widely studied in the social and behavioral sciences and, more recently, by the informatics and physical sciences communities [e.g., 4]. In particular, social scientists have constructed theories which qualitatively and quantitatively explain these processes and their dependence on social networks [e.g., 2-4, 6, 18, 36, 49-52]. One result of this research is a recognition that the process by which preferences and opinions of individuals become the collective outcome for a group can be complex and subtle, and thus challenging to model and predict. People arrive at their decisions by reacting individually to an environment consisting largely of others who are reacting likewise, and one consequence of this feedback dynamics is that the collective outcome can be quite different from one implied by a simple aggregation of individual preferences.

We model PEP in a manner which explicitly separates the individual, or "micro", dynamics from the collective dynamics. More specifically, we adopt a modeling framework consisting of three modeling scales:

- a *micro-scale*, for modeling the behavior of individuals;
- a *meso-scale*, which represents the interaction dynamics of individuals within the same network partition element (community or core/periphery);
- a *macro-scale*, which characterizes the interaction between partition elements.

We now derive a few properties of the multi-scale model. The micro-scale quantifies the way individuals combine their own inherent preferences regarding the available options with their observations of the behaviors of others to arrive at their chosen courses of action. Interestingly, the dependence of this decision-making process on the social network admits a straightforward characterization. Consider the common and important binary choice setting, in which N agents choose from a set $O = \{0,1\}$ of options based in part on the choices made by others. Let $o_i \in \{0,1\}$ denote the selection of agent i and $o = [o_1 \ldots o_N]^T \in O^N$ represent the vector of choices made by the group. It is reasonable to suppose that agent i chooses between the options probabil-istically according to some map $PO_i$: $A_i \times O^N \to [0,1]$, where $PO_i$ is the probability that agent i chooses option 1, $A_i$ measures i's inherent preference for option 1, and $PO_i$ is nondecreasing in $A_i$. In positive externality situations $PO_i$ should also be "nondecreasing in o" in some sense, and we now make this notion precise. (For notational simplicity in what follows we suppress the dependence of $PO_i$ on $A_i$.)

Because it is defined in such general terms it may appear that the map $PO_i$ could be a very complicated function of the choices of the other agents. In fact, Theorem 1 indicates that this map must be tractable.

**Theorem 1:** Given any $PO_i$ there exists a vector $w_i = [w_{i1} \ldots w_{iN}]^T \in \Re^N$, with $w_{ij} \geq 0$ and $\Sigma_j w_{ij} = b_i$, and a scalar function $r_i$: $[0, b_i] \to [0,1]$ such that $PO_i(o) = r_i(o^T w_i)$.

**Proof:** It is enough to prove that the $w_{ij}$ can be chosen so $o^T w_i$: $O^N \to [0, b_i]$ is injective, since then $r_i$ can be constructed to recover any $PO_i$. One such choice for $w_i$ is $w_i = [2^0\ 2^1 \ldots 2^{N-1}]^T$, as then $o^T w_i$ provides a unique (binary number) representation for each o. ∎

We call $r_i$ the *agent decision function* and $s_i = o^T w_i$ agent i's *social signal*, and interpret the $w_{ij}$ as defining a weighted social network for the group of N agents. Observe that Theorem 1 quantifies the way social influence is transmitted to an agent by her neighbors and highlights the importance of this signal in the decision-making process. The result also allows a simple characterization of positive externality agent behavior: for such behavior, $r_i$ is nondecreasing in $s_i$.

The micro-scale model structure allows PEP behaviors which appear to be distinct to be represented within a unified setting. For example, the basic model readily accommodates two of the most common sources of PEP: 1.) *utility-oriented externalities*, in which the utility or value of an option is a direct function of the number of others choosing it, and 2.) *information externalities*, which arise from inferences made by an individual about decision-relevant information possessed by others.

**Example A1.1: utility-oriented externalities.** Suppose each agent i has a utility function $u_i$: $O \times [0, b_i] \to \Re^+$ which depends explicitly on i's social signal $s_i$. The standard, albeit dated, example here is the fax machine, with the utility of owning a fax machine increasing with the number of others who own one. The key quantity considered by agent i when selecting between options 0 and 1 is the utility difference between the options, $\Delta u_i(s_i) = u_i(1,s_i) - u_i(0,s_i)$. In positive externality situations $\Delta u_i$ is increasing in $s_i$, and there exists a *threshold* social signal value s*, possibly with s* < 0 or s* > $b_i$, such that a utility maximizing agent will choose option 0 if $s_i$ < s* and option 1 if $s_i \geq$ s*.

**Example A1.2: information externalities.** Suppose the utility to agent i of each option is independent of the number of other agents choosing that option but there exists uncertainty regarding this utility. To be concrete, assume that agent i's utility depends on the "state of world" $w \in \{w_0, w_1\}$, so that $u_i = u_i(o_i,w)$, and there exists uncertainty regarding w. In this case, agent i may observe others' decisions in order to infer w and then choose the option which maximizes his utility for this world state (as when a tourist chooses a crowded restaurant over an empty one in an unfamiliar city). Consider, for instance, the decision of whether to adopt an innovation of uncertain quality, and let the world state $w_1$ signify that innovation quality is such that adopting maximizes utility. In this situation it is reasonable for agent i to maximize *expected* utility and choose the option (adopt or not) $o_i^* = \mathrm{argmax}_{o \in O} \Sigma_{w \in W} P(w \mid s_i) u_i(o_i,w)$. If agent i uses Bayesian inference to estimate $P(w_1 \mid s_i)$ then we have a positive externality decision process and there exists a threshold value $s^*$ for the social signal such that agent i will choose option 0 if $s_i < s^*$ and option 1 if $s_i \geq s^*$ [17].

It can be seen that in these examples, different positive externality "drivers" lead to equivalent (threshold) micro-scale models.

Taken together, the meso- and macro-scale components of the proposed modeling framework quantify the way agent decision functions interact to produce collective behavior at the population level. For convenience of exposition, in this Appendix we focus on network communities as the meso-scale structure of interest; however, all of the modeling results derived here also hold for the case of core-periphery structure . The role of the meso-scale model is to quantify and illuminate the manner in which agent decision functions interact *within* social network communities, while the macro-scale model characterizes the interactions of agents *between* communities. The primary assumption is that interactions between individuals within social network communities can be modeled as "fully-mixed" – all pairwise interactions between individuals within a network community are equally likely – while interactions between communities are constrained by the network defining the relationships between the communities. We argue below that this assumption is reasonable and useful.

One advantage of identifying a scale at which agent interaction is (approximately) homogeneous is that this enables the leveraging of an extensive literature on collective dynamics. To be concrete, we derive two examples. Consider first the social movement model proposed in [49,50]. In this model, each individual can be in one of three states: member (of the movement), potential member, and ex-member. Individuals interact in a fully-mixed way, with each interaction between a potential member and a member resulting in the potential member becoming a member with probability $\beta'$, and each interaction between a member and an ex-member resulting in the member becoming an ex-member with probability $\delta_1'$; members also "spontaneously" become ex-members with probability $\delta_2'$. The connection between this representation and standard epidemiological models [1] is clear.

Under the assumption of fully-mixed interactions at the meso-scale, standard manipulations yield the following representation for the social dynamics within network communities:

$$\Sigma_H: \begin{aligned} dP/dt &= -\beta PM - (\beta PM)^{1/2}\eta_1(t), \\ dM/dt &= \beta PM + (\beta PM)^{1/2}\eta_1(t) - \delta_1 ME - (\delta_1 ME)^{1/2}\eta_2(t) - \delta_2 M - (\delta_2 M)^{1/2}\eta_3(t), \\ dE/dt &= \delta_1 ME + (\delta_1 ME)^{1/2}\eta_2(t) + \delta_2 M + (\delta_2 M)^{1/2}\eta_3(t), \end{aligned}$$

where P, M, and E denote the fractions of potential members, members, and ex-members in the community population, $\beta$, $\delta_1$, and $\delta_2$ are nonnegative constants related to the probabilities $\beta'$, $\delta_1'$, and $\delta_2'$ defined above, and the $\eta_i(t)$ are appropriate random processes [e.g., 17]. The deterministic version of this basic model (i.e., with $\eta_1(t)=\eta_2(t)=\eta_3(t)\equiv 0$) is discussed by Hedstrom and co-authors in [49,50], and therefore we denote the model $\Sigma_H$. The deterministic version is shown in [49] to provide a useful description for the *local* growth of a real world social movement.

The second example incorporates the fact that innovations often have both enthusiasts and skeptics, each of whom may actively attempt to recruit the uncommitted. The model $\Sigma_H$ can be modified to account for this competition in recruitment:

$$\Sigma_B: \begin{aligned} dP/dt &= -\beta_1 PM_1 - \beta_2 PM_2, \\ dM_1/dt &= \beta_1 PM_1 - \delta_1 M_1, \\ dM_2/dt &= \beta_2 PM_2 - \delta_2 M_2, \\ dE/dt &= \delta_1 M_1 + \delta_2 M_2, \end{aligned}$$

where P and E denote the fractions of potential members and ex-members, as before, $M_1$ and $M_2$ are members of the competing groups or movements, and $\beta_1$, $\beta_2$, $\delta_1$, and $\delta_1$ are nonnegative constants. A model of this basic form is proposed in Bettencourt and coworkers in [51] and thus we label it $\Sigma_B$. The model can be fitted, with good agreement, to empirical data for the diffusion of Feynman diagrams (an innovation in physics) in the post World War II era [51]. Developing a stochastic version of $\Sigma_B$, analogous to the representation $\Sigma_H$, is straightforward [39].

The meso-scale model describes the way individual agent decision functions interact to produce collective behavior within social network communities. Individuals also interact with people from other communities, of course, and receive information from channels that transmit to many communities simultaneously (e.g., mass media). These inter-community interactions and "global" social signals are quantified at the macro-scale level of the multi-scale modeling framework. The basic idea is simple and natural: we model interdependence between social network communities with a graph $G_{sc} = \{V_{sc}, E_{sc}\}$, where $V_{sc}$ and $E_{sc}$ are the vertex and edge sets, respectively, $|V_{sc}| = K$, each vertex $v \in V_{sc}$ is a community, and each directed edge $e = (v,v') \in E_{sc}$ represents a potential inter-community interaction. More specifically, an edge $(v,v')$ indicates that an agent in community $v'$ can receive decision-relevant information from one in community

v. The way agents act upon this information is specified by their decision functions $r_i$. The broadcast of global social signals to individuals is modeled as a community-dependent input $u_v$ to each individual in community v. Thus $G_{sc}$ and the $u_v$ define the macro-scale model structure.

A key task in deriving a macro-scale model is specifying the topology of $G_{sc}$, as this graph encodes the social network structure for the phenomenon of interest. The most direct approach to constructing $G_{sc}$ is to infer communities directly from social network data, by partitioning the network so as to maximizing the graph modularity $Q_m$. The main challenge with this method for building social community graphs is obtaining the requisite social network data. While this task is certainly nontrivial, availability of such data has increased dramatically over the past decade. For instance, social relationships and interactions increasingly leave "fingerprints" in electronic databases (e.g., communication via email and cell phones, financial transactions), making convenient the acquisition, manipulation, storage, and analysis of these records [e.g., 4].

Alternatively, demographics data can sometimes be used to define both the communities themselves (e.g., families, physical neighborhoods) and their proximity. The basic idea is familiar: individuals belong to social groups, which in turn belong to "groups of groups", and so on, giving rise to a hierarchical organization of communities. For instance, in academics, research groups often belong to academic departments, which are organized into colleges, which in turn form universities, and so on. The proximity of two communities is specified by their relationship within the hierarchy, and this distance defines the likelihood that individuals from the two communities will interact. The probability of inter-community interaction, in turn, can be used to define the network community graph $G_{sc}$ [39].

**A1.2 S-HDS Model Formulation**

We now show that the stochastic hybrid dynamical system formalism provides a rigorous, tractable, and expressive framework within which to represent multi-scale social dynamics models. Consider the following

**Definition A1.1:** A *stochastic hybrid dynamical system* (S-HDS) is a feedback interconnection of a continuous-time, continuous state-dependent Markov chain $\{Q, \Lambda(x)\}$ and a collection of stochastic differential equations indexed by the Markov chain state q:

$$\Sigma_{\text{S-HDS}}: \quad \begin{aligned} &\{Q, \Lambda(x)\}, \\ &dx = f_q(x,p)dt + G_q(x,p)dw, \end{aligned}$$

where $q \in Q$ is the discrete state, $x \in X \subseteq \Re^n$ is the continuous state, $p \in \Re^p$ is a vector of system parameters, $\{f_q\}$ and $\{G_q\}$ are sets of vector and matrix fields characterizing the continuous system dynamics, w is an m-valued Weiner process, and $\Lambda(x)$ is the matrix of (x-dependent) Markov chain transition rates; the entries of $\Lambda(x)$ satisfy $\lambda_{qq'}(x) \geq 0$ if $q \neq q'$ and $\Sigma_{q'} \lambda_{qq'}(x) = 0$ $\forall q$, and are related to the standard Markov state transition probabilities as follows [e.g., 34]:

$$P\{q(t+\Delta) = q' | q(t) = q\} = \begin{cases} \lambda_{qq'}(x(t))\Delta + o(\Delta) & \text{if } q \neq q' \\ 1 + \lambda_{qq}(x(t))\Delta + o(\Delta) & \text{if } q = q'. \end{cases}$$

A general discussion of S-HDS theory and applications is beyond the scope of this paper and may be found in, for instance, [34] and the references therein.

We now develop an S-HDS representation for multi-scale social diffusion processes. It is assumed that:

- the social system consists of N individuals distributed over K network communities;
- individuals can influence each other via positive externalities;
- intra-community interactions are fully-mixed;
- inter-community interactions involve the (possibly temporary) migration of individuals from one community to another.

The phenomenon of interest is the diffusion of innovations, in which an innovation of some kind (e.g., a new technology or idea) is introduced into a social system, and individuals may learn about the innovation from others and decide to adopt it [e.g., 2]. By definition an innovation is "new", and therefore it is supposed that initially only a few of the network communities have been exposed to it. An important task in applications is to be able to characterize the likelihood that the innovation will spread to a significant fraction of the population [17].

We model social diffusion as follows:

**Definition A1.2:** The *multi-scale S-HDS diffusion model* is a tuple

$$\Sigma_{\text{S-HDS, diff}} = \{G_{sc}, Q \times X, \{f_q(x), G_q(x), H_q(x)\}_{q \in Q}, \text{Par}, W, U, \{Q, \Lambda(x)\}\}$$

where

- $G_{sc} = \{V_{sc}, E_{sc}\}$ is the social network community graph;
- $Q \times X$ is the system state set, with Q and $X \subseteq \Re^n$ denoting the (finite) discrete and (bounded) continuous state sets, respectively;
- $\{f_q(x), G_q(x), H_q(x)\}_{q \in Q}$, Par, W, U is the S-HDS continuous system, a family of stochastic differential equations which characterizes the intra-community dynamics via vector field/matrix families $\{f_q\}, \{G_q\}, \{H_q\}$, system parameter vector $p \in \text{Par} \subseteq \Re^p$, and system inputs $w \in W \subseteq \Re^m$, $u \in U \subseteq \Re^r$;
- $\{Q, \Lambda(x)\}$ is the S-HDS discrete system, a continuous-time Markov chain which defines inter-community interactions via state set Q and transition rate matrix $\Lambda(x)$.

The social community graph $G_{sc}$ defines the feasible community-community innovation diffusion pathways: if $(v,v') \notin E_{sc}$ then it is not possible for the innovation to spread directly from community v to community v'. The discrete state set $Q = \{0,1\}^K$ specifies which communities contain at least one adopter of the innovation by labeling such communities with a '1' (and a '0' otherwise). Thus, for example, state $q = [1\ 0\ 0\ \ldots]^T$ indicates that community 1 has at least one adopter, community 2 and 3 do not, and so on. The continuous state space X has coordinates $x_{ij} \in [0,1]$, where $x_{ij}$ is the ith state variable for the continuous system dynamics evolving in community j. For consistency we use the first coordinate for each community, $x_{1j}$, to refer to the fraction of adopters for that community. The continuous system dynamics is defined by a family of q-indexed stochastic differential equations $\{\Sigma_{cs,q}\}_{q \in Q}$, with

$$\Sigma_{cs,q}: \qquad dx = f_q(x,p)dt + G_q(x,p)dw + H_q(x,p)du,$$

where $w \in W$ is a standard Weiner process and $u \in U$ is the exogenous input. Ordinarily w is interpreted as a stochastic "disturbance", while u is employed to represent influences from "global" sources such as mass media. These dynamics quantify intra-community diffusion of the innovation of interest, for instance through models of the form $\Sigma_H$. The Markov chain matrix $\Lambda(x)$ specifies the transition rates for discrete state transitions $q \to q'$ and depends on both $G_{sc}$ and x (e.g., the rate at which community v will "infect" other communities depends upon the fraction of adopters in v). It is worth noting that the model $\Sigma_{S\text{-HDS, diff}}$ naturally accommodates both probabilistic (via w and the Markov chain dynamics) and set-bounded (through parameter set Par) uncertainty descriptions, as this expressiveness is desirable in applications.

### A1.3 A Simple Example

We now demonstrate the implementation of the proposed multi-scale S-HDS diffusion modeling framework, and illustrate its efficacy, through a simple example; a more complex example, with more interesting analytic goals, is investigated in Appendix Two below. Consider a social network consisting of two communities and a social movement process playing out on this network. We construct the social network using the method given in [52]. Briefly, a collection of N vertices is divided into two communities of equal size, denoted L and R (for 'left' and 'right', see Figure 6). For all vertex pairs, if both vertices belong to the same community then an edge is placed between them with probability $p_i$, and if the vertices belong to different communities then they are connected with probability $p_e < p_i$. Increasing the ratio $p_i / p_e$ makes the resulting network more "community-like" by increasing the relative intra-community edge density. Figure 6 shows two small example networks built in this way, with the network on the left corresponding to a larger $p_i / p_e$ ratio.

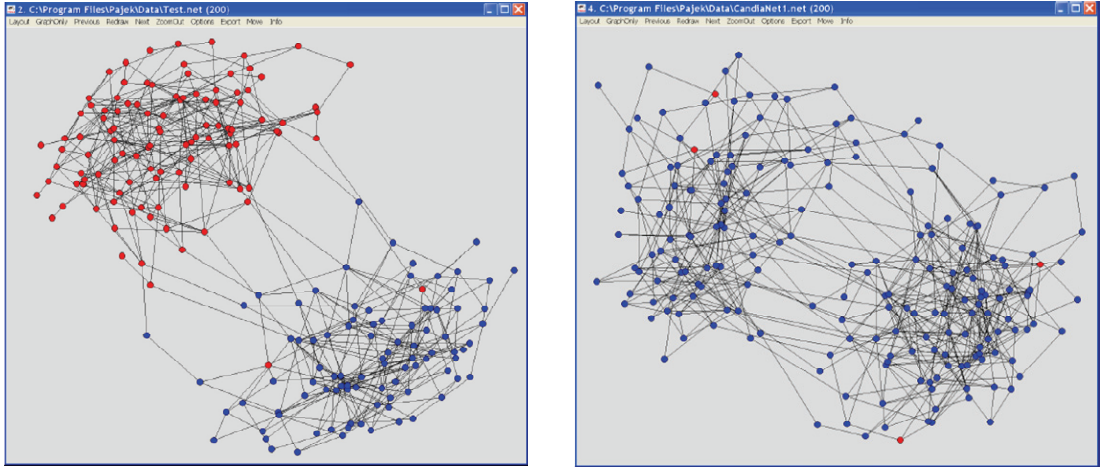

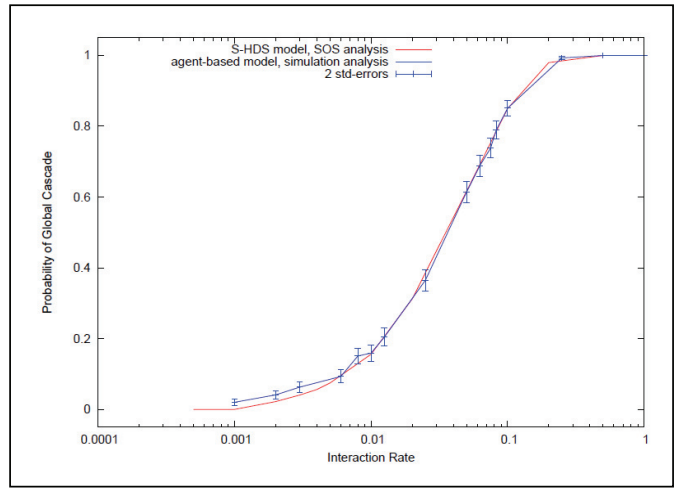

**Figure 6.** Sample results for ABM/S-HDS comparison study. The visualization at top is a cartoon of the basic setup, in which an innovation is introduced into one of the two network communities comprising a social system; possible outcomes include diffusion of the innovation throughout the community initially "infected" (left network, blue vertices are in state M) or across both communities (right network). The plot at bottom shows the probability of "global" diffusion as a function of inter-community interaction intensity for the models $\Sigma_{ABM}$ (blue curve) and $\Sigma_{S\text{-}HDS,\,diff}$ (red curve).

The social movement dynamics evolving on this network is a "network version" of the model proposed in [49]. Thus each individual can be in one of three states – member, potential member, and ex-member – and individuals can change states in one of three ways: 1.) members persuade potential members to whom they are linked to become members with probability $\beta'$, 2.) ex-members likewise influence neighboring members to become ex-members with probability $\delta_1'$, and 3.) members can spontaneously become ex-members with probability $\delta_2'$. For convenience of reference this "agent-based" system representation is denoted $\Sigma_{ABM}$.

It is straightforward to derive an S-HDS version of the social movement model $\Sigma_{ABM}$. Consider the diffusion model $\Sigma_{\text{S-HDS, diff}} = \{G_{sc}, Q \times X, \{f_q(x), G_q(x), H_q(x)\}_{q \in Q}, \text{Par}, W, U, \{Q, \Lambda(x)\}\}$ specified in Definition A1.2. Note first that in this case the social network community graph $G_{sc}$ is very simple, consisting of two vertices corresponding to communities L and R and an undirected edge connecting them. The continuous system state is $x = [P_L \ M_L \ P_R \ M_R]^T \in X$, where the subscripts indicate communities (note that the concentrations of ex-members, $E_L$ and $E_R$, are not independent states because the total concentration sums to one on each community). We approximate the agent-based social movement dynamics *within* each network community with the fully-mixed model $\Sigma_H$, that is, with a set of stochastic differential equations governing the evolution of the concentrations of members M and potential members P.

It can be seen that $\Sigma_H$ together with the preceding discussion defines the model components $X$, $\{f_q(x), G_q(x), H_q(x)\}_{q \in Q}$, Par, W, U that make up the continuous system portion of $\Sigma_{\text{S-HDS, diff}}$. Thus all that remains is to specify the discrete system $\{Q, \Lambda(x)\}$. The discrete state set $Q = \{00, 10, 01, 11\}$ indicates which communities contain at least one movement member, so that for instance state $q = 10$ indicates that community L has at least one member and community R has no members. The Markov chain matrix $\Lambda(x)$ specifies the transition rates for discrete state transitions $q \to q'$. These rates depend on the continuous system state x because the likelihood that one community will "infect" the other depends upon the current concentrations of members, potential members, and ex-members in that community.

We examine the utility of the S-HDS social diffusion model constructed above by using this model to estimate the probability that a small set of "seed" members introduced into community L will lead to the movement growing and eventually propagating to community R. Because the model $\Sigma_{\text{S-HDS, diff}}$ is derived from $\Sigma_{ABM}$, $\Sigma_{ABM}$ is taken to be ground truth and $\Sigma_{\text{S-HDS, diff}}$ is deemed a useful approximation if the cascade probability estimates obtained using the S-HDS representation are in good agreement with those computed based on $\Sigma_{ABM}$. The following parameter values are chosen for $\Sigma_{ABM}$: $N = 2000$, $\beta' = 0.5$, $\delta_1' = 0.01$, $\delta_2' = 0.1$ (the results reported are not sensitive to variation in these values). We build 50 random realizations of the social network for each of 15 $p_i / p_e$ ratios. The values for $p_i / p_e$ are selected to generate a collection of 15 network sets whose topologies interpolate smoothly between networks with essentially disconnected communities (large $p_i / p_e$) and networks whose two communities are tightly coupled (small $p_i / p_e$). A "global" cascade is said to occur if an initial seed set of five movement members in community R, chosen at random, results in the diffusion of the movement to community L. The probability of global cascade at a given $p_i / p_e$ ratio is computed by running 20 simulations on each of the 50 social network realizations associated with that $p_i / p_e$, and counting up those for which the innovation propagates to community L. The results of this simulation study are presented in the plot at the bottom of Figure 6, with the blue curve showing the probability estimates as a function of $p_i / p_e$ ratio and the error bars corresponding to $\pm 2$ standard errors.

We now investigate the efficacy of the S-HDS social diffusion model by using this model to estimate the probability of global cascade. The social diffusion model $\Sigma_{\text{S-HDS, diff}}$ is instantiated to be equivalent to the agent-based representation $\Sigma_{\text{ABM}}$ described above. Note that, in particular, there are no free parameters available to permit the response of $\Sigma_{\text{S-HDS, diff}}$ to be "tuned" to match $\Sigma_{\text{ABM}}$. For instance, the $\Sigma_{\text{ABM}}$ parameters $\beta'$, $\delta_1'$, $\delta_2'$ uniquely define $\Sigma_{\text{S-HDS, diff}}$ parameters $\beta$, $\delta_1$, $\delta_2$, and specifying values for the $p_i / p_e$ ratios gives corresponding values for the S-HDS transition matrices $\Lambda(x)$ (to within a single "offset" parameter, see [39]). A Matlab program implementing the resulting model $\Sigma_{\text{S-HDS, diff}}$ is given in [39].

In order to compute the probability of global cascade using the S-HDS model $\Sigma_{\text{S-HDS, diff}}$, we employ the "altitude function" method described in Appendix Two below. This method calculates provably-correct upper bounds on the probability of the social movement propagating to community L. The results of this analysis are given at the plot of the bottom of Figure 6 (red curve). Observe that the global cascade probability estimates obtained using the two models $\Sigma_{\text{ABM}}$ and $\Sigma_{\text{S-HDS, diff}}$ are in close agreement. As it is challenging to model "discontinuous" phenomena such as diffusion across social network communities, this agreement represents important evidence that the S-HDS provides a useful characterization of social diffusion on networks.

While the models $\Sigma_{\text{ABM}}$ and $\Sigma_{\text{S-HDS, diff}}$ generate similar results in this example, the S-HDS representation is much more efficient computationally. For instance, estimating the desired global cascade probabilities using the S-HDS model requires less than one percent of the computer time needed to obtain these estimates with the equivalent agent-based model. Moreover, this difference on efficiency increases with network size, which is important because realistic social networks have hundreds or thousands of communities rather than just two. This computational tractability hints at a more general, and more significant, mathematical tractability enjoyed by the S-HDS framework, a property we now leverage to develop a rigorous predictive analysis methodology for social diffusion events.

**A2. Appendix Two: Predictive Analysis**

In this Appendix we formulate the predictive analysis problem in terms of reachability assessment, show that these reachability questions can be addressed through an "altitude function" analysis *without computing system trajectories*, and apply this theoretical framework to demonstrate that predictability of a broad class of social diffusion models depends crucially upon the meso-scale topological structures of the underlying networks. For convenience of exposition, in this Appendix we focus on network communities as a representative meso-scale structure; how-ever, all results derived here are also applicable to the more general case in which the "network partition" (see Section 2.2) includes both community and core-periphery structures.

**A2.1 Predictive Analysis as Reachability Assessment**

We propose that accurate prediction requires careful consideration of the interplay between the intrinsics of a process and the social dynamics which are its realization. We therefore adopt an inherently dynamical approach to predictive analysis: given a social process, a set of measurables, and the behavior of interest, we formulate prediction problems as questions about the reachability properties of the system. Toward that end, the behavior about which predictions are to be made is used to define the system *state space subsets of interest* (SSI), while the particular set of candidate measurables under consideration allows identification of the *candidate starting set* (CSS), that is, the set of states and system parameter values which represent initializations that are equivalent under the assumed observational capability. This setup permits predictability assessment, and the related task of identifying useful measurables, to be performed in a systematic manner. Roughly speaking, the proposed approach to predictability assessment involves determining how probable it is to reach the SSI from a CSS and deciding if these reachability properties are compatible with the prediction goals. If a system's reachability characteristics are incompatible with the given prediction question – if, say, "hit" and "flop" in a cultural market are both likely to be reached from the CSS – then the prediction objectives should be refined in some way. Possible refinements include relaxing the level of detail to be predicted or introducing additional measurables.

We now make these notions more precise. Consider the multi-scale S-HDS social diffusion model $\Sigma_{\text{S-HDS, diff}}$ specified in Definition A1.2. Let $P_0$ be a subset of the parameter set Par and $X_0$, $X_{s1}$, $X_{s2}$ be subsets of the (bounded) continuous system state space X. Suppose $X_0 \times P_0$ and $\{X_{s1}, X_{s2}\}$ are the CSS and SSI, respectively, corresponding to the prediction question. Let a specification $\delta > 0$ be given for the minimum acceptable level of variation in system behavior relative to $\{X_{s1}, X_{s2}\}$. Consider the following

**Definition A2.1:** A situation is *eventual state (ES) predictable* if $|\gamma_1 - \gamma_2| > \delta$, where $\gamma_1$ and $\gamma_2$ are the probabilities of $\Sigma_{\text{S-HDS, diff}}$ reaching $X_{s1}$ and $X_{s2}$, respectively, and is *ES unpredictable* otherwise.

Note that in ES predictability problems it is expected that the two sets $\{X_{s1}, X_{s2}\}$ represent qualitatively different system behaviors (e.g., hit and flop in a cultural market), so that if the probabilities of reaching each from $X_0 \times P_0$ are similar then system behavior is unpredictable in a sense that is meaningful for many applications. Other useful forms of predictability are defined and investigated in [39].

The notion of predictability forms the basis for our definition of useful measurables:

**Definition A2.2:** Let the components of the vectors $(x_0, p_0) \in X_0 \times P_0$ which comprise the CSS be denoted $x_0 = [x_{01} \ldots x_{0n}]^T$ and $p_0 = [p_{01} \ldots p_{0p}]^T$. The *measurables with most predictive power* are those state variables $x_{0j}$ and/or parameters $p_{0k}$ for which predictability is most sensitive.

Intuitively, those measurables for which predictability is most sensitive are likely to be the ones that can most dramatically affect the predictability of a given problem. Note that we do not specify a particular measure of sensitivity to be used when identifying measurables with maximum predictive power, as such considerations are ordinarily application-dependent (see [39] for some useful specifications). Definitions A2.1 and A2.2 focus on the role played by *initial* states in the predictability of social processes. In some cases it is useful to expand this formulation to allow consideration of states other than initial states. For instance, we show in [18] that very early time series are often predictive for PEP, suggesting that it can be valuable to consider initial state *trajectory segments*, rather than just initial states, when assessing predictability. This extension can be naturally accomplished by redefining the CSS, for instance by augmenting the state space X with an explicit time coordinate [18].

We now turn our attention to the "early warning" problem.

**Definition A2.3:** Let the event of interest be specified in terms of $\Sigma_{\text{S-HDS, diff}}$ reaching or escaping some SSI $X_s$, and suppose a warning signal is to be issued only if the probability of event occurrence exceeds some specified threshold $\alpha$. *Reach warning analysis* involves identifying a state set $X_w$, where $X_s \subseteq X_w$ necessarily, with the property that if the system trajectory enters $X_w$ then the probability that $\Sigma_{\text{S-HDS, diff}}$ will eventually reach $X_s$ is at least $\alpha$. Analogously, *escape warning analysis* involves identifying a state set $X_w$, where $X \setminus X_w \subseteq X_s$ necessarily, with the property that if the system trajectory enters $X_w$ then the probability that $\Sigma_{\text{S-HDS, diff}}$ will eventually escape from $X_s$ is at least $\alpha$.

## A2.2 Stochastic Reachability Assessment

The previous section formulates predictive analysis problems as reachability questions. Here we show that these reachability questions can be addressed through an "altitude function" analysis, in which we seek a scalar function of the system state that permits conclusions to be made regarding reachability *without computing system trajectories*. We refer to these as altitude functions to provide an intuitive sense of their analytic role: if some measure of "altitude" is low on the CSS and high on an SSI, and if the expected rate of change of altitude along system trajectories is nonincreasing, then it is unlikely for trajectories to reach this SSI from the CSS.

Consider the S-HDS social diffusion model $\Sigma_{\text{S-HDS, diff}}$ evolving on a bounded state space $Q \times X$. We quantify the uncertainty associated with $\Sigma_{\text{S-HDS, diff}}$ by specifying bounds on the possible values for some system parameters and perturbations and giving probabilistic descriptions for other uncertain system elements and disturbances. Given this representation, it is natural to seek a probabilistic assessment of system reachability.

We begin with an investigation of probabilistic reachability on *infinite* time horizons. The following "supermartingale lemma" is proved in [53] and is instrumental in our development:

**Lemma SM:** Consider a stochastic process $\Sigma_s$ with bounded state space X, and let $\underline{x}(t)$ denote the "stopped" process associated with $\Sigma_s$ (i.e., $\underline{x}(t)$ is the trajectory of $\Sigma_s$ which starts at $x_0$ and is stopped if it encounters the boundary of X). If $A(\underline{x}(t))$ is a nonnegative supermartingale then for any $x_0$ and $\lambda > 0$

$$P\{\sup A(\underline{x}(t)) \geq \lambda \mid \underline{x}(0) = x_0\} \leq A(x_0) / \lambda.$$

Denote by $X_0 \subseteq X$ and $X_s \subseteq X$ the initial state set and SSI, respectively, for the continuous system component of $\Sigma_{\text{S-HDS, diff}}$, and assume that X and the parameter set Par $\subseteq \Re^p$ are both bounded. Thus, for instance, the SSI is a subset of the continuous system state space X alone; this is typically the case in applications and is easily extended if necessary. We are now in a position to state our first stochastic reachability result:

**Theorem 2:** $\gamma$ is an upper bound on the probability of trajectories of $\Sigma_{\text{S-HDS, diff}}$ reaching $X_s$ from $X_0$, while remaining in Q × X, if there is a family of differentiable functions $\{A_q(x)\}_{q \in Q}$ such that

- $A_q(x) \leq \gamma \ \forall x \in X_0, \ \forall q \in Q$;
- $A_q(x) \geq 1 \ \forall x \in X_s, \ \forall q \in Q$;
- $A_q(x) \geq 0 \ \forall x \in X, \ \forall q \in Q$;
- $(\partial A_q/\partial x)(f_q + H_q u) + (1/2) \, \text{tr} \, [G_q^T (\partial^2 A_q/\partial x^2) G_q] + \Sigma_{q' \in Q} \lambda_{qq'} A_{q'} \leq 0 \ \forall x \in X, \ \forall q \in Q, \ \forall u \in U, \ \forall p \in \text{Par}.$

**Proof:** Note first that $BA_q(x) = (\partial A_q/\partial x)(f_q + H_q u) + (1/2) \, \text{tr} \, [G_q^T (\partial^2 A_q/\partial x^2) G_q] + \Sigma_{q' \in Q} \lambda_{qq'} A_{q'}$ is the infinitesimal generator for $\Sigma_{\text{S-HDS, diff}}$, and therefore quantifies the evolution of the expectation of $A_q(x)$ [53,34]. As a consequence, the third and fourth conditions of the theorem imply that $A(q(t),x(t))$ is a nonnegative supermartingale [53]. Thus, from Lemma SM, we can conclude that $P\{x(t) \in X_s \text{ for some t}\} \leq P\{\sup A(q(t),x(t)) \geq 1 \mid x(0)=x_0\} \leq A(q,x_0) \leq \gamma \ \forall x_0 \in X_0, \ \forall q \in Q, \ \forall u \in U, \ \forall p \in \text{Par}.$ ∎

The preceding result characterizes reachability of S-HDS on infinite time horizons. In some situations, including important applications involving social systems, it is of interest to study system behavior on *finite* time horizons. The following result is useful for such analysis:

**Theorem 3:** $\gamma$ is an upper bound on the probability of trajectories of $\Sigma_{\text{S-HDS, diff}}$ reaching $X_s$ from $X_0$ during time interval $[0,T]$, while remaining in $Q \times X$, if there exists a family of differentiable functions $\{A_q(x,t)\}_{q \in Q}$ such that

- $A_q(x,t) \leq \gamma \;\; \forall (x,t) \in X_0 \times 0, \; \forall q \in Q$;
- $A_q(x,t) \geq 1 \;\; \forall (x,t) \in X_s \times [0,T], \; \forall q \in Q$;
- $A_q(x,t) \geq 0 \;\; \forall (x,t) \in X \times \mathfrak{R}^+, \; \forall q \in Q$;
- $BA_q(x,t) \leq 0 \;\; \forall (x,t) \in X \times \mathfrak{R}^+, \; \forall q \in Q, \; \forall u \in U, \; \forall p \in \text{Par}$.

**Proof:** The proof follows immediately from that of Theorem 2 once it is observed that $P\{\underline{x}(t) \in X_s$ for some $t \in [0,T]\} = P\{(\underline{x}(t),t) \in X_s \times [0,T]\}$. ∎

The idea for the proof of Theorem 3 was suggested in [54].

Having formulated predictability assessment for social processes in terms of system reachability and presented a new theoretical methodology for assessing reachability, we are now in a position to give our approach to deciding predictability. Observe first that Theorems 2 and 3 are of direct practical interest only if it is possible to efficiently compute a tight probability bound $\gamma$ and associated altitude function $A(x)$ which satisfy the theorem conditions. Toward that end, observe that the theorems specify *convex* conditions to be satisfied by altitude functions: if $A_1$ and $A_2$ satisfy the theorem conditions then any convex combination of $A_1$ and $A_2$ will also satisfy the conditions. Thus the search for altitude functions can be formulated as a convex programming problem [55]. Moreover, if the system of interest admits a polynomial description (e.g., the system vector and matrix fields are polynomials) and we search to polynomial altitude functions, then the search can be carried out using sum-of-squares (SOS) optimization [56].

SOS optimization is a convex relaxation framework based on SOS decomposition of the relevant polynomials and semidefinite programming. SOS relaxation involves replacing the nonnegative and nonpositive conditions to be satisfied by the altitude functions with SOS conditions. For example, the conditions for $A_q(x)$ given in Theorem 2 can be relaxed as follows:

$$
\begin{aligned}
A(x) \leq \gamma \;\; \forall x \in X_0 \;\; &\rightarrow \;\; \gamma - A(x) - \lambda_0^T(x) g_0(x) \text{ is SOS} \\
A(x) \geq 1 \;\; \forall x \in X_s \;\; &\rightarrow \;\; A(x) - 1 - \lambda_s^T(x) g_s(x) \text{ is SOS} \\
A(x) \geq 0 \;\; \forall x \in X \;\; &\rightarrow \;\; A(x) - \lambda_{X1}^T(x) g_{X1}(x) \text{ is SOS} \\
BA(x) \leq 0 \;\; \forall x \in X, \forall p \in \text{Par} \;\; &\rightarrow \;\; -BA(x) - \lambda_{X2}^T(x) g_{X2}(x) - \lambda_P^T(p) g_P(p) \text{ is SOS}
\end{aligned}
$$

where the entries of the vector functions $\lambda_0, \lambda_s, \lambda_{X1}, \lambda_{X2}, \lambda_P$ are SOS, the vector functions $g_0, g_s, g_{X1}, g_{X2}, g_P$ satisfy $g_*(\cdot) \geq 0$ (entry-wise) whenever $x \in X_*$ or $p \in \text{Par}$, respectively, and we assume $|Q| = 1$ for notational convenience. The conditions on $A_q(x,t)$ specified in Theorem 3 can be relaxed in exactly the same manner. The relaxed SOS conditions are clearly sufficient and in practice are typically not overly-conservative [56,39].

Once the set of conditions to be satisfied by A(x) are relaxed in this way, SOS programming can be used to compute $\gamma_{min}$, the minimum value for the probability bound $\gamma$, and A(x), the associated altitude function which certifies the correctness of this bound. Software for solving SOS programs is available as the third-party Matlab toolbox SOSTOOLS [56], and example SOS programs are given in [39]. Importantly, the approach is tractable: for fixed polynomial degrees, the computational complexity of the associated SOS program grows polynomially in the dimension of the continuous state space, the cardinality of the discrete state set, and the dimension of the parameter space.

For completeness, we outline an algorithm for computing the pair $(\gamma_{min}, A(x))$:

**Algorithm A2.1: altitude functions via SOS programming** (outline)

1. Parameterize A as $A(x) = \Sigma_k c_k a_k(x)$, where $\{a_1, \ldots, a_B\}$ are monomials up to a desired degree bound and $\{c_1, \ldots, c_B\}$ are to-be-determined coefficients.
2. Relax all A(x) criteria in the relevant theorem to SOS conditions.
3. Formulate an SOS program with decision variables $\gamma$, $\{c_1, \ldots, c_B\}$, where the desired bound on altitude function polynomial degree is reflected in the specification of the set $\{c_1, \ldots, c_B\}$. Compute the minimum probability bound $\gamma_{min}$ and values for the coefficients $\{c_1, \ldots, c_B\}$ that define A(x) using SOSTOOLS.

It is emphasized that, although the computation of $(\gamma_{min}, A(x))$ is performed numerically, the resulting function A(x) is guaranteed to satisfy the conditions of the relevant theorem and therefore represents a proof of the correctness of the probability upper bound $\gamma_{min}$. Note also that the probability estimate is obtained without computing system trajectories, and is valid for entire sets of initial states $X_0$, parameter values Par, and exogenous inputs U.

Having given a method for efficiently computing pairs $(\gamma_{min}, A(x))$, and thereby characterizing reachability, we are now in a position to sketch an algorithm for assessing ES predictability:

**Algorithm A2.2: ES predictability** (outline)

Given: social diffusion process of interest is $\Sigma_{\text{S-HDS, diff}}$, CSS = $X_0 \times P_0$, SSI = $\{X_{s1}, X_{s2}\}$, and minimum acceptable level of variation = $\delta$.

Procedure:
1. compute (upper bound for) probability $\gamma_1$ of $\Sigma_{\text{S-HDS, diff}}$ reaching $X_{s1}$ from $X_0 \times P_0$;
2. compute (upper bound for) probability $\gamma_2$ of $\Sigma_{\text{S-HDS, diff}}$ reaching $X_{s2}$ from $X_0 \times P_0$;
3. if $|\gamma_1 - \gamma_2| > \delta$ then problem is ES predictable, else problem is ES unpredictable.

Note: $\gamma_1, \gamma_2$ can be computed using Theorem 2 (infinite time horizon) or Theorem 3 (finite time horizon) together with Algorithm 3.1 and SOSTOOLS [56].

## A2.3 Application to Social Diffusion

The theoretical framework developed in the preceding sections is now used, in combination with empirically-grounded models for social diffusion [e.g., 17,49-51], to demonstrate that predictability of this class of diffusion models depends crucially upon network community structure. We investigate the following predictability question: Is the diffusion of social movements and mobilizations ES predictable and, if so, which measurable quantities have predictive power?

We adopt a specific version of the S-HDS social diffusion model proposed in Definition 2.2:

$$\Sigma_{\text{S-HDS, diff}} = \{G_{sc}, Q \times X, \{f_q(x), G_q(x)\}_{q \in Q}, \text{Par}, W, \{Q, \Lambda(x)\}\}$$

where

- the social network community graph $G_{sc}$ consists of K communities (so $|V_{sc}| = K$), connected together with an Erdos-Renyi random graph topology, with community size drawn from a power law distribution [36];
- each continuous system $\Sigma_{cs, q}$: $dx = f_q(x,p)dt + G_q(x,p)dw$, $q \in Q$, is given by the meso-scale social movement model $\Sigma_H$ or $\Sigma_B$ with appropriate parameter vector p and system "noise" w;
- the discrete system $\{Q, \Lambda(x)\}$ is a Markov chain that defines inter-community interactions in the manner described in Definition A1.2.

A Matlab instantiation of this S-HDS diffusion model is given in [39] and is available upon request. The behavior of the model can be shown to be consistent with empirical observations of several historical social movements (e.g., various movements in Sweden) [39].

In order to assess ES predictability, SSI = $\{X_{s1}, X_{s2}\}$ is defined so that $X_{s1}, X_{s2}$ are state sets corresponding to *global* (affecting a significant fraction of the population) and *local* (remaining confined to a small fraction of the population) movement events, respectively. We then employ Algorithm A2.2 iteratively to search for a definition for CSS = $X_0 \times P_0$ which ensures that the probabilities of reaching $X_{s1}$ and $X_{s2}$ from $X_0 \times P_0$ are sufficiently different to yield an ES predictable situation. We use two models of the form $\Sigma_{\text{S-HDS, diff}}$ for this analysis, corresponding to the two definitions for the continuous system $\Sigma_H$ and $\Sigma_B$. Each model is composed of K = 10 communities connected together with an Erdos-Renyi random graph topology. (Using different realizations of the Erdos-Renyi random graph does not affect the conclusions reported below.)

ES predictability analysis yields two main results. First, both the intra-community and inter-community dynamics exhibit *threshold* behavior: small changes in either the intra-community "infectivity" or inter-community interaction rate around their threshold values lead to large variations in the probability that the movement will propagate "globally". More quantitatively, for the diffusion model $\Sigma_{\text{S-HDS, diff}}$ with continuous system dynamics $\Sigma_H$, threshold behavior is obtained when varying 1.) the generalized reproduction number $R = \beta / \delta_2$ and 2.) the rate $\lambda$ at which in-ter-community interactions between individuals take place. Thus in order for a social movement to propagate to a significant fraction of the population, the threshold conditions R≥1 and λ≥$λ_0$ must be satisfied simultaneously. An analogous conclusion holds when $Σ_H$ is replaced with the diffusion model $Σ_B$ in the S-HDS representation. This finding is reminiscent of and extends well-known results for epidemic thresholds in disease propagation models [1].

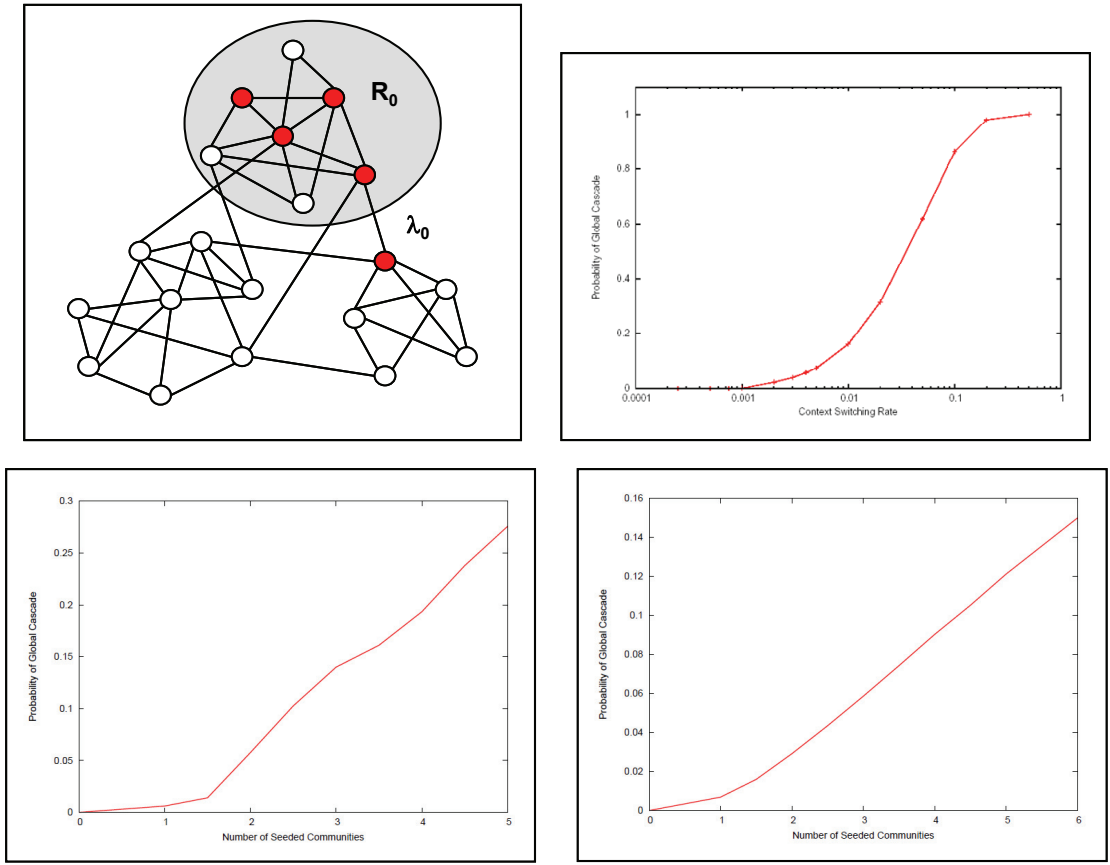

**Figure 7.** Sample results from social diffusion predictability study. Cartoon at top left illustrates the setup for the inter-community interaction study, highlighting the parameter values $R_0$=1 and $λ_0$ which quantify intra-and inter-community propagation thresholds; plot at top right shows classic threshold dependence of global propagation probability on inter-community interaction intensity λ. Plots in bottom row depict the way global propagation probability increases with the number of communities across which a fixed set of innovating seeds are distributed (plots at left and right show cascade probabilities for multi-scale models possessing $Σ_H$ and $Σ_B$ meso-scale dynamics, respectively).

This threshold behavior is illustrated in the plot at the top right of Figure 7, which shows the way probability of global propagation increases with inter-community interaction rate when the intra-community diffusion is sufficiently infective (i.e., R≥1). The probabilities which make up this plot represents provably-correct (upper bound) estimates computed using Theorem 2 and Algorithm A2.1. A similar threshold response is observed when varying intra-community infectivity R, provided the inter-community interaction rate satisfies $\lambda \geq \lambda_0$. Importantly, the inter-community interaction threshold $\lambda_0$ is seen to be quite small, indicating that even a few links between network communities enables rapid diffusion of the movement to otherwise disparate regions of the social network. This result suggests that a useful predictor of movement activity in a given community is the level of movement activity among that community's neighbors in $G_{sc}$.

The second main ES predictability result characterizes the way probability of global propagation varies with the number of network communities across which a *fixed* set of "seed" movement members is distributed. To quantify this dependence, the social movement model $\Sigma_{S\text{-HDS, diff}}$ is initialized so that a small fraction of individuals in the population are movement members and the remainder of the population consists solely of potential members. We then vary the way this initial seed set of movement members is distributed across the K network communities, at one extreme assigning all seeds to the same community and at the other spreading the seeds uniformly over all K communities. For each distribution of seed movement members, the probability of global movement propagation is computed using Theorem 2 and Algorithm A2.1. Other than initialization strategy, the model is specified exactly as in the preceding analysis.

The results of this portion of the ES predictability assessment are summarized in the two plots at the bottom of Figure 7. It is seen that for both choices of meso-scale social movement dynamics, $\Sigma_H$ and $\Sigma_B$, the probability of global movement propagation increases approximately linearly with the number of network communities across which the fixed set of seed members is distributed (here the number of initial members is set to one percent of the total population).